\documentclass[11pt]{article}
\pdfoutput=1
\usepackage{mathptmx}
\usepackage{amsmath,amssymb,amsthm}
\usepackage{authblk}
\usepackage[final]{graphicx}
\usepackage{color}
\usepackage[margin=0.75in]{geometry}
\usepackage{chngcntr}
\usepackage{indentfirst}
\usepackage{enumitem}
\usepackage{hyperref}
\usepackage{float}
\usepackage{bbm}
\usepackage{natbib}
\usepackage{appendix}
\usepackage{apxproof}

\DeclareMathOperator*{\argmax}{arg\,max}

\title{Distributionally Robust XVA via Wasserstein Distance \\
Part 2: Wrong Way Funding Risk}
\author{Derek Singh, \, Shuzhong Zhang}
\affil{Department of Industrial and Systems Engineering, University of Minnesota \\ singh644@umn.edu, \, zhangs@umn.edu}
\date{\vspace{-5ex}}

\theoremstyle{case}

\newtheorem{remark}{Remark}
\newtheoremrep{prop}{Proposition}[section]
\newtheoremrep{theorem}{Theorem}[section]
\newtheoremrep{lemma}[theorem]{Lemma}
\newtheoremrep{conj}[theorem]{Conjecture}
\newtheoremrep{corollary}{Corollary}[prop]

\providecommand{\keywords}[1]
{
  \small	
  \textbf{\textit{Keywords---}} #1
}

\usepackage[section]{placeins}

\begin{document}

\maketitle

\begin{abstract}
This paper investigates calculations of robust funding valuation adjustment (FVA) for over the counter (OTC) derivatives under distributional uncertainty using Wasserstein distance as the ambiguity measure. Wrong way funding risk can be characterized via the robust FVA formulation. The simpler dual formulation of the robust FVA optimization is derived. Next, some computational experiments are conducted to measure the additional FVA charge due to distributional uncertainty under a variety of portfolio and market configurations. Finally some suggestions for future work, such as robust capital valuation adjustment (KVA) and margin valuation adjustment (MVA), are discussed.
\end{abstract}


\section{Introduction and Overview}
\subsection{Financial Markets Context and Background}
\indent Funding valuation adjustment (FVA) represents the impact on portfolio market value due to funding exposures for the hedge on uncollateralized derivatives. It represents the market value of funding exposure risk. Funding cost adjustment (FCA) can be represented mathematically as an integral of discounted expected positive exposure times funding cost (incremental) conditional on joint counterparty and firm survival. FCA arises for a positive portfolio exposure since this implies a negative hedge exposure which leads to a funding cost for collateral posted. The market valuation is a function of joint counterparty and firm credit risk, the underlying (market) risk factors that drive the portfolio valuation (and hence positive exposure) as well as funding cost, as well as the correlations between these market risk factors and the credit risk curves for a given portfolio. FCA is typically measured and reported at the funding netting set level. \par
The ``other side" of FCA is funding benefit adjustment (FBA). This represents the funding benefit to the firm, for interest income proceeds on received collateral posted against counterparty exposure on the hedge, as measured by discounted expected negative exposure times funding benefit conditional on joint counterparty and firm survival. As above, the market valuation is a function of counterparty and firm credit risk, underlying market risk factors that drive portfolio valuation and funding benefit, and the correlations. FBA can be represented mathematically as an integral of discounted negative exposure times funding benefit  conditional on joint counterparty and firm survival. FBA is typically measured at the funding netting set level. \par
(Bilateral) FVA represents the dual impact on portfolio market value due to both funding cost and funding benefit exhibited over the portfolio lifetime. FVA can be represented mathematically as the difference (or sum) of two integrals: (i) discounted expected positive exposure times funding cost conditional on joint counterparty and firm survival (ii) discounted expected negative exposure times funding benefit conditional on joint counterparty and firm survival. FVA is typically measured and reported at the netting set level for a given firm. \par
As mentioned in Part 1 \citep{Singh19xva1}, U.S. regulatory authorities, the Federal Reserve and Office of the Comptroller of the Currency (OCC), periodically assess national banks' compliance with the Market Risk Capital Rule (MRR). Counterparty credit risk (CCR) and funding risk (FR) metrics are key metrics used to evaluate bank risk profiles due to OTC derivatives. Basel Committee on Banking Supervision, through Basel III, has developed criteria to quantify capital charges due to CCR. According to International Swap Dealers Association (ISDA) the current OTC derivatives notional outstanding is over 500 trillion. Consequently the CCR and FR exposures (due to uncollateralized or partially collateralized hedges) inherent in the OTC derivatives market represent significant market risk exposures. This motivates the concepts of worst case FVA and wrong way risk (WWR) and the impact of uncertainty in probability distribution on FR and FVA. It is these considerations that motivate this line of research. \citep {Ramzi19}, \citep{Hajjaji15} \par
An outline of this paper is as follows. Section 1 represents an introduction and overview of FVA and wrong way funding risk. Section 2 develops the main theoretical results of the paper and provides proof sketches. Section 3 conducts a computational study of WWR for a representative set of derivative instruments, portfolios, and market environments. Section 4 discusses the conclusions and suggestions for future research. All detailed proofs of propositions, corollaries, and theorems are deferred to the Appendix. \par

\subsection{Literature Review}
\begin{remark}
The authors are not aware of any substantial research that has been done on the topic of worst case FVA. The discussion below pertains to literature regarding worst case CVA.
\end{remark}
In the past few years research has been done to investigate the effect of distributional uncertainty on CVA. 
\citet{brigo2013counterparty} explicitly incorporate correlation into the stochastic processes driving the market risk and credit default factors. They measure the effect of dependency structure (and hence wrong way risk) on CVA for a variety of asset classes. \citet{glasserman15} bound the effect of wrong way risk on CVA. Their approach considers a discrete setting and formulates worst case CVA as the solution to a worst case linear program subject to certain constraints, where the dependency structure is allowed to vary. They introduce a penalty term, measured via Kullback-Leibler (KL) divergence, to control the degree of wrong way risk. Memartoluie, in his PhD thesis, uses an ordered scenario copula methodology to quantify worst case CVA \citep{Amir17}. 
For worst case correlations set to one, he finds results that are comparable to the method of Glasserman and Yang. \par

Lagrangian duality results were recently (and independently) developed by \citet{blanchetFirst} and \citet{Gao16}. These results hold under mild assumptions.
The main innovation in our work is to apply these recent results to worst case FVA using Wasserstein distance as the ambiguity measure. Furthermore, analytical expressions are derived for the solutions to the inner and outer convex optimization problems. A computational study shows the material impact of distributional uncertainty on worst case FVA and illustrates the risk profile.\par

\subsection{Notation and Definitions}
\subsubsection{FCA}
Notation and core definitions for FCA problem setup follow conventions in \citet {glasserman15} and \citet{Lichters16}.
Those for the robust FCA problem formulation follow conventions in \citet {blanchetMV}.
FCA measures expected funding cost over the lifetime of the portfolio. Let $V^{+}(t)$ denote the positive portfolio exposure at time $t$. The problem setup here assumes a fixed set of observation dates, $0 = t_0 < t_1 < \cdots < t_n = T$.
Let $X^+$ denote the vector of discounted positive exposures and $Y_C$ and $Y_F$ denote the vectors of counterparty and firm survival indicators. Further, let $Y_{CF}$ denote the Hadamard product $Y_C \odot Y_F$ which represents the vector of joint survival indicators. To incorporate funding, let $Z^+$ denote the vector of funding costs incurred on exposures $X^+$. Let $(z^+_i,y^{cf}_i)$ denote realizations of $(Z^+,Y_{CF})$ along sample paths for $i = \{1,2, \ldots ,N\}$.\par


The FCA associated with funding costs $Z^{+}$ and joint survival indicator $\mathbbm{1}_{\{\tau_C > t\} \cap \{\tau_F > t\}}$ is \citep {Lichters16}, \citep {green2015xva}
\begin{equation*}
\text{FCA} = \int_0^T \mathbb{E}[Z^+(t) \mathbbm{1}_{\{\tau_C > t\} \cap \{\tau_F > t\}} ] dt.
\end{equation*}

\noindent The pair of vectors $(Z^+,Y_{CF}) \in (\mathbb{R}^n_{+} \times B^1_n)$ is
\begin{equation*}
Z^+ = (f_c(t_0,t_1)X^{+}(t_1),\ldots,f_c(t_{n-1},t_n)X^{+}(t_n)) \quad \text{and} \quad Y_{CF} = (\mathbbm{1}_{\{\tau_C > t_1\} \cap \{\tau_F > t_1\}},\dots,\mathbbm{1}_{\{\tau_C > t_n\} \cap \{\tau_F > t_n\}}).
\end{equation*}
Here $B^1_n$ denotes the set of survival time vectors: binary vectors of ones and zeros with $n$ components, and at most one block of ones followed by a complementary block of zeros.
The empirical measure, $P_N$, is
\begin{equation*}
P_N(dz) = \frac{1}{N} \sum_{i=1}^{N} \mathbbm{1}_{(z^+_i,y^{cf}_i)} (dz).
\end{equation*}
Under the empirical measure, $P_N$, FCA is an expectation of an inner product
\begin{equation*}
\text{FCA} = \mathbb{E}^{P_N} [\langle Z^+,Y_{CF} \rangle].
\end{equation*}
In the context of this work, the uncertainty set for probability measures is
\begin{equation*}
\mathcal{U}_{\delta_1}(P_N) = \{P: D_c(P,P_N) \leq \delta_1\}
\end{equation*}
where $D_c$ is the optimal transport cost or Wasserstein discrepancy for cost function $c$ \citep {blanchetMV}.
For convenience the definition for $D_c$ is
\begin{equation*}
D_c(P,P') = \inf \{ \mathbb{E}^\pi[c(A,B)]: \pi \in \mathcal{P}(\mathbb{R}^d \times \mathbb{R}^d), \pi_A = P, \pi_B = P' \}
\end{equation*}
where $\mathcal{P}$ denotes the space of Borel probability measures and $\pi_A$ and $\pi_B$ denote the distributions of $A$ and $B$. 
Here $A$ denotes $(Z^+_A,Y_A) \in (\mathbb{R}^n_{+} \times B^1_n)$ and $B$ denotes $(Z^+_B,Y_B) \in (\mathbb{R}^n_{+} \times B^1_n)$ respectively.
This work uses the cost function $c_{S_1}$ where
\begin{equation*}
c_{S_1}((u,v),(z,y)) = S_1 \langle v-y,v-y \rangle + \langle u-z,u-z \rangle.
\end{equation*}
The scale factor $S_1 > 0$ is used to compensate for different domains: $(u,v) \in (\mathbb{R}^n_{+} \times B^1_n), (z,y) \in (\mathbb{R}^n_{+} \times B^1_n)$.

\subsubsection{FBA}
Notation and core definitions for FBA problem setup follow conventions in \citet {glasserman15} and \citet {Lichters16}.
Those for the robust FBA problem formulation follow conventions in \citet {blanchetMV}.
FBA measures expected funding benefit over the lifetime of the portfolio. Let $V^{-}(t)$ denote the negative portfolio exposure at time $t$. The problem setup here assumes a fixed set of observation dates, $0 = t_0 < t_1 < \cdots < t_n = T$.
Let $X^-$ denote the vector of discounted negative exposures and $Y_C$ and $Y_F$ denote the vectors of counterparty and firm survival indicators. Further, let $Y_{CF}$ denote the Hadamard product $Y_C \odot Y_F$ which represents the vector of joint survival indicators. To incorporate funding, let $Z^-$ denote the vector of funding benfits incurred on exposures $X^-$. Let $(z^-_i,y^{cf}_i)$ denote realizations of $(Z^-,Y_{CF})$ along sample paths for $i = \{1,2, \ldots ,N\}$.\par


The FBA associated with funding benefits $Z^{-}$ and joint survival indicator $\mathbbm{1}_{\{\tau_C > t\} \cap \{\tau_F > t\}}$ is \citep {Lichters16}, \citep {green2015xva}
\begin{equation*}
\text{FBA} = \int_0^T \mathbb{E}[Z^-(t) \mathbbm{1}_{\{\tau_C > t\} \cap \{\tau_F > t\}} ] dt.
\end{equation*}

\noindent The pair of vectors $(Z^-,Y_{CF}) \in (\mathbb{R}^n_{-} \times B^1_n)$ is
\begin{equation*}
Z^- = (f_b(t_0,t_1)X^{-}(t_1),\ldots,f_b(t_{n-1},t_n)X^{-}(t_n)) \quad \text{and} \quad Y_{CF} = (\mathbbm{1}_{\{\tau_C > t_1\} \cap \{\tau_F > t_1\}},\dots,\mathbbm{1}_{\{\tau_C > t_n\} \cap \{\tau_F > t_n\}}).
\end{equation*}
Here $B^1_n$ denotes the set of survival time vectors: binary vectors of ones and zeros with $n$ components, and at most one block of ones followed by a complementary block of zeros.
The empirical measure, $Q_N$, can be written as
\begin{equation*}
Q_N(dz) = \frac{1}{N} \sum_{i=1}^{N} \mathbbm{1}_{(z^-_i,y^{cf}_i)} (dz).
\end{equation*}
Under the empirical measure, $Q_N$, FBA is an expectation of an inner product
\begin{equation*}
\text{FBA} = \mathbb{E}^{Q_N} [\langle Z^-,Y_{CF} \rangle].
\end{equation*}
In the context of this work, the uncertainty set for probability measures is
\begin{equation*}
\mathcal{U}_{\delta_2}(Q_N) = \{Q: D_c(Q,Q_N) \leq \delta_2\}
\end{equation*}
where $D_c$ is the optimal transport cost or Wasserstein discrepancy for cost function $c$ \citep {blanchetMV}.
For convenience the definition for $D_c$ is
\begin{equation*}
D_c(Q,Q') = \inf \{ \mathbb{E}^\pi[c(A,B)]: \pi \in \mathcal{P}(\mathbb{R}^d \times \mathbb{R}^d), \pi_A = Q, \pi_B = Q' \}
\end{equation*}
where $\mathcal{P}$ denotes the space of Borel probability measures and $\pi_A$ and $\pi_B$ denote the distributions of $A$ and $B$. 
Here $A$ denotes $(Z^-_A,Y_A) \in (\mathbb{R}^n_{-} \times B^1_n)$ and $B$ denotes $(Z^-_B,Y_B) \in (\mathbb{R}^n_{-} \times B^1_n)$ respectively.
This work uses the cost function $c_{S_2}$ where
\begin{equation*}
c_{S_2}((u,v),(x,y)) = S_2 \langle v-y,v-y \rangle + \langle u-z,u-z \rangle.
\end{equation*}
The scale factor $S_2 > 0$ is used to compensate for different domains: $(u,v) \in (\mathbb{R}^n_{-} \times B^1_n), (z,y) \in (\mathbb{R}^n_{-} \times B^1_n)$.

\subsubsection{FVA}

Notation and core definitions for (bilateral) FVA problem setup incorporate those above for FCA and FBA.
FVA measures expected funding costs and benefits over portfolio lifetime. Let $V^{+}(t)$ denote the positive portfolio exposure at time $t$.  Let $V^{-}(t)$ denote the negative portfolio exposure at time $t$. The problem setup here assumes a fixed set of observation dates, $0 = t_0 < t_1 < \cdots < t_n = T$.
Let $X^+$ denote the vector of discounted positive exposures and $Y_C$ denote the vector of counterparty survival indicators. 
Let $X^-$ denote the vector of discounted negative exposures and $Y_F$ denote the vector of firm survival indicators. 
Further, let $Y_{CF}$ denote the Hadamard product $Y_C \odot Y_F$ which represents the vector of joint survival indicators. To incorporate funding, let $Z^+$ denote the vector of funding costs incurred on exposures $X^+$. And similarly for $Z^-$ with respect to exposures $X^-$.
Due to the linkage between $Z^+$ and $Z^-$, one can write $Z = Z^{+} + Z^-$ and decompose sample realizations of $Z$ into $Z^+$ and $Z^-$ accordingly.
Therefore, let $(z_i,y^{cf}_i)$ denote realizations of $(Z,Y_{CF})$ along sample paths for  $i = \{1,2, \ldots ,N\}$. 
The relation $z_i = z^+_i + z^-_i$ can be used to decompose $z_i$ into its positive and negative exposures respectively. \par


The FVA associated with funding costs $Z(t)$, joint survival indicator $\mathbbm{1}_{\{\tau_C > t\} \cap \{\tau_F > t\}}$ is \citep {Lichters16}, \citep {green2015xva}
\begin{equation*}
\text{FVA} = \text{FCA} + \text{FBA} = \int_0^T \mathbb{E}[Z^+(t) \mathbbm{1}_{\{\tau_C > t\} \cap \{\tau_F > t\}} ] dt + \int_0^T \mathbb{E}[Z^-(t) \mathbbm{1}_{\{\tau_C > t\} \cap \{\tau_F > t\}} ] dt =  \int_0^T \mathbb{E}[Z(t) \mathbbm{1}_{\{\tau_C > t\} \cap \{\tau_F > t\}} ] dt.
\end{equation*}
\noindent The pair of vectors $(Z,Y_{CF}) \in (\mathbb{R}^n \times B^1_n)$ is
\begin{equation*}
Z = (Z^{+}(t_1)+Z^{-}(t_1),\ldots,Z^{+}(t_n)+Z^{-}(t_n)) \quad \text{and} \quad Y_{CF} = (\mathbbm{1}_{\{\tau_C > t_1\} \cap \{\tau_F > t_1\}},\dots,\mathbbm{1}_{\{\tau_C > t_n\} \cap \{\tau_F > t_n\}}),
\end{equation*}
and the pair of vectors $(Z^+,Z^-) \in (\mathbb{R}^n_{+} \times \mathbb{R}^n_{-})$ is
\begin{equation*}
Z^+ = (f_c(t_0,t_1)X^{+}(t_1),\ldots,f_c(t_{n-1},t_n)X^{+}(t_n)) \quad \text{and} \quad Z^- = (f_b(t_0,t_1)X^{-}(t_1),\ldots,f_b(t_{n-1},t_n)X^{-}(t_n)).
\end{equation*}
Here $B^1_n$ denotes the set of survival time vectors: binary vectors of ones and zeros with $n$ components, and at most one block of ones followed by a complementary block of zeros.
The empirical measure, $\Phi_N$, is
\begin{equation*}
\Phi_N(dz) = \frac{1}{N} \sum_{i=1}^{N} \mathbbm{1}_{(z_i,y^{cf})} (dz).
\end{equation*}
Under the empirical measure, $\Phi_N$, FVA is a sum of expectations of inner products
\begin{equation*}
\text{FVA} = \mathbb{E}^{\Phi_N} [\langle Z^+,Y_{CF} \rangle] +  \mathbb{E}^{\Phi_N} [\langle Z^-,Y_{CF} \rangle] = \mathbb{E}^{\Phi_N} [\langle Z,Y_{CF} \rangle].
\end{equation*}
In the context of this work, the uncertainty set for probability measures is
\begin{equation*}
\mathcal{U}_{\delta_3}(\Phi_N) = \{P: D_c(\Phi,\Phi_N) \leq \delta_3\}
\end{equation*}
where $D_c$ is the optimal transport cost or Wasserstein discrepancy for cost function $c$ \citep {blanchetMV}.
For convenience the definition for $D_c$ is
\begin{equation*}
D_c(\Phi,\Phi') = \inf \{ \mathbb{E}^\pi[c(A,B)]: \pi \in \mathcal{P}(\mathbb{R}^d \times \mathbb{R}^d), \pi_A = \Phi, \pi_B = \Phi' \}
\end{equation*}
where $\mathcal{P}$ denotes the space of Borel probability measures and $\pi_A$ and $\pi_B$ denote the distributions of $A$ and $B$. 
Here $A$ denotes $(Z_A,Y_A) \in (\mathbb{R}^n \times B^1_n)$ and $B$ denotes $(Z_B,Y_B) \in (\mathbb{R}^n \times B^1_n)$ respectively.
This work uses the cost function $c_{S_3}$ where
\begin{equation*}
c_{S_3}((u,v),(z,y)) = S_3 \langle v-y,v-y \rangle + \langle u-z,u-z \rangle.
\end{equation*}
The scale factor $S_3 > 0$ is used to compensate for different domains: $(u,v) \in (\mathbb{R}^n \times B^1_n), (z,y) \in (\mathbb{R}^n \times B^1_n)$. 

\begingroup
\setlength{\parindent}{0pt}
\section{Theory: Robust FVA and Wrong Way Funding Risk}
\subsection{FCA}
\subsubsection{Inner Optimization Problem}
The robust FCA can be written as
\begin{equation*}\label{eqn:primal}
\sup_{P \in \mathcal{U}_{\delta_1}(P_N)} \mathbb{E}^P [\langle Z^+, Y_{CF} \rangle]  \tag{P1}.
\end{equation*}

Now use recent duality results, noting the inner product $\langle \: ; \rangle$ satisfies the upper semicontinuous condition of the Lagrangian duality theorem, and cost function $c_S$ satisfies the non-negative lower semicontinuous condition (see \citet{blanchetFirst} Assumptions 1 \& 2, \citet{Gao16}). Hence the dual problem (to sup above) can be written as
\begin{equation*}\label{eqn:dual}
 \inf_{\gamma \geq 0} \: H(\gamma) := \bigg[ \gamma \delta_1 + \frac{1}{N} \sum_{i=1}^{N} \Psi_\gamma(z^+_i,y^{cf}_i) \bigg]  \tag{D1}
\end{equation*}
where 
\begin{align*}
\Psi_\gamma(z^+_i,y^{cf}_i)  = \sup_{u \in \mathbb{R}^n_{+}, v \in {B^1_n}} [  \langle u,v \rangle - \gamma c_{S_1}((u,v),(z^+_i,y^{cf}_i)) ] \,\,
									= \sup_{u \in \mathbb{R}^n_{+}, v \in {B^1_n}} [  \langle u, v \rangle - \gamma( \langle u-z^+_i, u-z^+_i \rangle + S_1 \langle v-y^{cf}_i, v-y^{cf}_i \rangle ) ].
\end{align*}


Now apply change of variables $w_1 = (u-z^+_i)$ and $w_2 = (v-y^{cf}_i)$ to get
\begin{equation*}
\Psi_\gamma(z^+_i,y^{cf}_i)  = \sup_{w_1 \geq -z^+_i, w_2 \in {B^2_n}} [ \langle w_1+z^+_i, w_2+y^{cf}_i \rangle - \gamma ( \langle w_1, w_1 \rangle + S_1 \langle w_2, w_2 \rangle ) ]
\end{equation*}
where $B^2_n$ denotes the set of ternary vectors of ones, zeros, and minus ones with $n$ components, and at most one block of ones or minus ones. 
Note that $\sup_{w_1} [ \,\, ; \, ]$ is attained for $w_1^{*} \in \mathbb{R}^n_{+}$ (as will become evident in the proof) hence it suffices to consider this space for $w_1$.
It turns out that $\Psi_\gamma$ can be expressed as original FCA plus the pointwise max of $(n+1)$ convex functions. The degenerate case $l=0$ is just a line of negative slope. The other $n$ cases are a hyperbola plus a line of negative slope. $\Psi_\gamma$ quantifies the adversarial move in FCA across both time and spatial dimensions while accounting for the cost via the $K$ terms.

\begin{proprep}
Let \, $\Psi_\gamma(z^+_i,y^{cf}_i) = \, \langle z^+_i,y^{cf}_i \rangle + \big[  \frac{l^*}{4 \gamma} + (\sum_{k=1}^{l^*}z^+_{ik} - \sum_{k=1}^{\|y^{cf}_i\|_1} z^+_{ik}) - \gamma S_1 K \big]$\\
where $l^{*} = \argmax_{l \geq 0} [ \frac{l}{4 \gamma} + \sum_{k=1}^{l} z^+_{ik} - \gamma S_1 K]$ and $l = \|w_2 + y^{cf}_i \|_1 \geq 0, \:  l \in \mathbb{Z^+}$.
Also $\|y^{cf}_i\|_1 \in \mathbb{Z^+}$, and  $K = | l - \| y^{cf}_i \|_1 | = \| w_2 \|_1 \geq 0, K \in \mathbb{Z^+}$.
Once $l^*$ is selected, $K := | l^* - \| y^{cf}_i \|_1 | = \| w_2^* \|_1$. 
Alternatively, 
$\Psi_\gamma(z^+_i,y^{cf}_i) = \, \langle z^+_i,y^{cf}_i \rangle + \bigvee_{l=0}^n  h_\gamma(l)$ for \\ $h_\gamma(l) := \big[  \frac{l}{4 \gamma} + (\sum_{k=1}^{l}z^+_{ik} - \sum_{k=1}^{\|y^{cf}_i\|_1} z^+_{ik}) - \gamma S_1 K \big]$.
Finally, note $\bigvee_{l=0}^n  h_\gamma(l)$ denotes $\max_{l \in \{0,\dots,n\}} h_\gamma(l)$.

\end{proprep}
\begin{proofsketch}
This result follows from jointly maximizing the adversarial funding exposure $w_1$ and the survival time index $w_2$.  The structure of $B^2_n$ allows us to decouple this joint maximization and find the critical point to maximize the quadratic in $w_1$ and write down the condition to select the optimal survival time index $l^*$. Finally, consider the two cases $w_2 = 0$ and $w_2 \neq 0$ and take the max to arrive at the solution. The $K$ terms represent the cost associated with the worst case.
\end{proofsketch}

\begin{appendixproof}
\begingroup
\setlength{\parindent}{0pt}

The particular structure of $B^1_n$ and $B^2_n$ will be exploited to evaluate the $\sup$ above.
The analysis proceeds by considering different cases for optimal values $(w_1^{*}, w_2^{*})$.\\
$\mathbf{Case \, 1}$ \quad
Suppose $w_2^{*} = 0 \implies  l = \| y^{cf}_i \|_1$. Then\\ 
\begin{equation*}
\Psi_\gamma(z^+_i,y^{cf}_i)  = \, \langle z^+_i,y^{cf}_i \rangle + \sup_{w_1 \in \mathbb{R}^n_{+}} [  \langle w_1, y^{cf}_i \rangle - \gamma \langle w_1, w_1 \rangle ].
\end{equation*}
Applying the Cauchy-Schwarz Inequality gives
\begin{equation*}
\Psi_\gamma(z^+_i,y^{cf}_i)  = \, \langle z^+_i,y^{cf}_i \rangle + \sup_{\| w_1 \|} [ \| w_1\| \| y^{cf}_i \| - \gamma \| w_1 \|^2 ].
\end{equation*}
Evaluating the critical point $\|w_1^{*}\| = \frac{\|y^{cf}_i\|}{2\gamma} \in \mathbb{R}_{+}$ for the quadratic gives
\begin{equation*}
\Psi_\gamma(z^+_i,y^{cf}_i)  = \, \langle z^+_i,y^{cf}_i \rangle + \frac{ \| y^{cf}_i \|^2}{4\gamma} = \langle z^+_i,y^{cf}_i \rangle + \frac{ \| y^{cf}_i \|_1}{4\gamma}.
\end{equation*}

$\mathbf{Case \, 2}$ \quad
Now consider $w_2^{*} \neq 0 \implies l \neq \| y^{cf}_i \|_1 $. \\
Observe for $l = \|w_2 + y^{cf}_i \|_1 \geq 0$, \\
\begin{equation*}
\langle w_1+z^+_i, w_2 + y_i^{cf} \rangle = \sum_{k=1}^{l} (w_{1k} + z^+_{ik}).
\end{equation*}
The structure of finite set $B^2_n$ implies
\begin{equation*}
\Psi_\gamma(z^+_i,y^{cf}_i) = \sup_{w_1 \in \mathbb{R}^n_{+}, l \in \{0,\ldots,n\}, l \neq  \| y^{cf}_i \|_1} [ \sum_{k=1}^{l} (w_{1k} + z^+_{ik})  - \gamma ( \langle w_1, w_1 \rangle + S_1 K ) ].
\end{equation*}

Again, using that $B^2_n$ is a finite set, one can write
\begin{equation*}
\Psi_\gamma(z^+_i,y^{cf}_i) = \max_{ l \in \{0,\ldots,n\}, l \neq  \| y^{cf}_i \|_1} \sup_{w_1 \in \mathbb{R}^n_{+}} [ \sum_{k=1}^{l} (w_{1k} + z^+_{ik}) - \gamma ( \langle w_1, w_1 \rangle + S_1 K ) ].
\end{equation*}
Observing that only the first $l$ components of $w_1 \text{ inside the sup are positive gives } \forall k \in \{1,\ldots,l\}$
\begin{equation*}
\sup_{w_1 \in \mathbb{R}^n_{+}} [ \sum_{k=1}^{l} (w_{1k}) - \gamma \langle w_1, w_1 \rangle  ]  = l \times \sup_{w_{1k} \in \mathbb{R}_{+}} [ w_{1k} - \gamma ( w_{1k} )^2 ].
\end{equation*}
Evaluating at the critical point $w^{*}_{1k} = \frac{1}{2\gamma} \in \mathbb{R}_{+}$ for the above quadratic gives
\begin{equation*}
\sup_{w_{1k} \in \mathbb{R}_{+}} [ w_{1k} - \gamma (w_{1k}^2 ) ] = \frac{1}{4\gamma}.
\end{equation*}
Therefore one can write
\begin{equation*}
\Psi_\gamma(z^+_i,y^{cf}_i) = \max_{l \in \{0,\ldots,n\}, l \neq  \| y^{cf}_i \|_1} [ \frac{l}{4\gamma} + \sum_{k=1}^{l} (z^+_{ik}) - \gamma S_1 K ].
\end{equation*}

Furthermore, $l^{*}$ is determined as
\begin{equation*}
l^{*} = \argmax_{l \in \{0,\dots,n\}, l \neq  \| y^{cf}_i \|_1} [ \frac{l}{4\gamma} + \sum_{k=1}^{l} (z^+_{ik}) - \gamma S_1 K ].
\end{equation*}

Substituting back into expression for $\Psi_\gamma$ gives
\begin{equation*}
\Psi_\gamma(z^+_i,y^{cf}_i) = \, \langle z^+_i,y^{cf}_i \rangle + \bigg[  \frac{l^*}{4 \gamma} +  \bigg( \sum_{k=1}^{l^*} z^+_{ik}  - \sum_{k=1}^{\|y^{cf}_i\|_1} z^+_{ik} \bigg)  - \gamma S_1 K \bigg].
\end{equation*}

Finally, taking the max values for $\Psi_\gamma$ over cases $w_2^{*} = 0$ and $w_2^{*} \neq 0$ gives
\begin{equation*}
\Psi_\gamma(z^+_i,y^{cf}_i) = \, \langle z^+_i,y^{cf}_i \rangle + \bigg[\frac{ \| y^{cf}_i \|_1}{4 \gamma}\bigg] \vee  \bigg[  \frac{l^*}{4 \gamma} + \bigg( \sum_{k=1}^{l^*} z^+_{ik}  - \sum_{k=1}^{\|y^{cf}_i\|_1} z^+_{ik} \bigg) - \gamma S_1 K \bigg].
\end{equation*}
Observe that for $l^* = \| y^{cf}_i \|_1$, the last term in brackets $[ \, ; ]$ above evaluates to $\big[\frac{ \| y^{cf}_i \|_1}{4 \gamma}\big]$.
Therefore let $l^*$ be determined as
\begin{equation*}
l^{*} = \argmax_{l \in \{0,\dots,n\}} [ \frac{l}{4\gamma} + \sum_{k=1}^{l} (z^+_{ik}) - \gamma S_1 K ]
\end{equation*}
and write
\begin{equation*}
\Psi_\gamma(z^+_i,y^{cf}_i) = \, \langle z^+_i,y^{cf}_i \rangle + \bigg[  \frac{l^*}{4 \gamma} + \bigg(\sum_{k=1}^{l^*}z^+_{ik} - \sum_{k=1}^{\|y^{cf}_i\|_1} z^+_{ik}\bigg) - \gamma S_1 K \bigg].
\end{equation*}

Alternatively, one can write
\begin{equation*}
\Psi_\gamma(z^+_i,y^{cf}_i) = \, \langle z^+_i,y^{cf}_i \rangle + \bigvee_{l=0}^n  \bigg[  \frac{l}{4 \gamma} + \bigg(\sum_{k=1}^{l}z^+_{ik} - \sum_{k=1}^{\|y^{cf}_i\|_1} z^+_{ik}\bigg) - \gamma S_1 K \bigg].
\end{equation*}.
\endgroup
\end{appendixproof}

\subsubsection{Outer Optimization Problem}
The goal now is to evaluate
\begin{equation*}\label{eqn:dual1}
 \inf_{\gamma \geq 0} \:  H(\gamma) := \bigg[ \gamma \delta_1 + \frac{1}{N} \sum_{i=1}^{N} \Psi_\gamma(z^+_i,y^{cf}_i) \bigg]  
\end{equation*}
where 
\begin{equation*}
\Psi_\gamma(z^+_i,y^{cf}_i)  = \, \langle z^+_i,y^{cf}_i \rangle + \bigvee_{l=0}^n h_\gamma(l) \:\: \text{where} \:\:  h_\gamma(l) := \big[  \frac{l}{4 \gamma} + \big(\sum_{k=1}^{l}z^+_{ik} - \sum_{k=1}^{\|y^{cf}_i\|_1} z^+_{ik}\big) - \gamma S_1 K \big].
\end{equation*}
The convexity of the objective function $H(\gamma)$ simplifies the task of solving this optimization problem. The first order optimality condition suffices. As $\Psi_\gamma$ and hence $H(\gamma)$ may have non-differentiable kinks due to the max functions, $\vee$, we characterize the optimality condition via subgradients. In particular, we look for $\gamma^{*} \geq 0$ such that $0 \in \partial H(\gamma^{*})$. Inspection of the asymptotic properties of $\Psi_\gamma$ and its subgradients reveals that $\partial H(\gamma)$ will cross zero (as $\gamma$ sweeps from $0$ to $\infty$) and hence $\gamma^{*} \geq 0$.

\begin{proprep}
Let $\gamma^{*} \in \,  \left\{ \gamma \geq 0: 0 \in \partial H(\gamma) \right\} \\ where \,\,
 \partial \Psi_\gamma = \mathbf{Conv} \cup \{ \partial h_\gamma(l) \: | \:  \langle z^+_i,y^{cf}_i \rangle + h_\gamma(l) = \Psi_\gamma ; \, l \in \{0,\dots,n\} \}$ 
and $\partial H(\gamma) = \delta_1 + \frac{1}{N} \sum_{i=1}^N \partial \Psi_\gamma$.
\end{proprep}
\begin{proofsketch}
This follows from application of standard properties of subgradients as well as inspection of the asymptotic properties of $\Psi_\gamma$ and $\partial \Psi_\gamma$. For $\gamma$ sufficiently small, $\Psi_\gamma$ has a large positive value and $\partial \Psi_\gamma$ has a large negative derivative. For $\gamma$ sufficiently large, for optimal $l^*$, either $l^* = 0 \implies 0 \in \partial \Psi_\gamma$ or $l^* = \| y^{cf}_i \|_1 > 0 \implies \partial \Psi_\gamma$ approaches zero $\implies \partial H(\gamma)$ crosses zero.
\end{proofsketch}
\begin{appendixproof}
\begingroup
\setlength{\parindent}{0pt}

This follows from standard application of properties of convex functions and subgradients. 
First note that function $h_\gamma$ is convex in $\gamma$ since (for fixed $l$) it is the sum of a hyperbola plus a constant plus a negative linear term. 
So then $\Psi_\gamma$ is convex since it is the pointwise max of a finite set of convex functions plus a constant. 
Using properties of subgradients, one can write $\partial \Psi_\gamma = \mathbf{Conv} \cup \{ \partial h_\gamma(l) \: | \: \langle z^+_i,y^{cf}_i \rangle + h_\gamma(l) = \Psi_\gamma ; \, l \in \{0,\dots,n\}\}$.
Furthermore $H(\gamma)$ is convex in $\gamma$ since it is a linear term plus a sum of convex functions, so one can write 
$\gamma^{*} \in \,  \{ \gamma : 0 \in \partial H(\gamma) \}$ and it follows that $\partial H(\gamma) = \delta_1 + \frac{1}{N} \sum_{i=1}^N \partial \Psi_\gamma$.
Finally, we argue that $\gamma^{*} \geq 0$. 
For $\gamma > 0$ sufficiently small, $\exists \, z  < -\delta_1$ such that $z \in \partial \Psi_\gamma$ and for $\gamma > 0$ sufficiently large, $\exists \, z > -\delta_1$ such that $z \in \partial \Psi_\gamma$. 
To elaborate, for $\gamma > 0$ sufficiently large, $\| y^{cf}_i \|_1 > 0 \implies l^* = \| y^{cf}_i \|_1 \implies K = 0 \implies \exists \, z > -\delta_1$ such that $z \in \partial \Psi_\gamma$.
To elaborate, for $\gamma > 0$ sufficiently large, $\| y^{cf}_i \|_1 = 0 \implies l^* = 0 \implies K = 0, \Psi_\gamma = 0, 0 = z > -\delta_1$ such that $z \in \partial \Psi_\gamma$.
Hence we deduce $\partial H(\gamma)$ crosses zero ( as $\gamma$ sweeps from $0$ to $\infty$ ). 
\endgroup
\end{appendixproof}

Putting together the results of these two propositions, we arrive at our first theorem.

\begin{theoremrep}
The primal problem \ref{eqn:primal} has solution
$\big[ \gamma^{*} \delta_1 + \frac{1}{N} \sum_{i=1}^{N} \Psi_{\gamma^{*}}(z^+_i,y^{cf}_i) \big]$ \\
where $\gamma^{*} \in \,  \left\{ \gamma \geq 0 : 0 \in \partial H(\gamma) \right\}$ and
$\Psi_{\gamma^{*}}(z^+_i,y^{cf}_i) = \, \langle z^+_i,y^{cf}_i \rangle + \bigvee_{l=0}^n h_{\gamma^{*}}(l) \:\: \text{for} \:\:  h_{\gamma^{*}}(l) := \big[  \frac{l}{4 \gamma^{*}} + (\sum_{k=1}^{l}z^+_{ik} - \sum_{k=1}^{\|y^{cf}_i\|_1} z^+_{ik}) - \gamma^{*} S_1 K \big]$.
Expressed in terms of original FCA, this says
\begin{equation*}
\sup_{P \in \mathcal{U}_{\delta_1}(P_N)} \mathbb{E}^P [\langle Z^+,Y_{CF} \rangle] = \mathbb{E}^{P_N} [\langle Z^+,Y_{CF} \rangle] + \gamma^{*} \delta_1 +   \mathbb{E}^{P_N} \big[ \bigvee_{l=0}^n \frac{l}{4 \gamma^{*}} + \big(\sum_{k=1}^{l}Z^+_k - \sum_{k=1}^{\|Y_{CF}\|_1} Z^+_k \big) - \gamma^{*} S_1 K  \big]
\end{equation*}
where the additional terms represent a penalty due to uncertainty in probability distribution.
\end{theoremrep}
\begin{proofsketch}
This follows directly from the previous two propositions.
\end{proofsketch}
\begin{appendixproof}
This follows by direct substitution of $\gamma^{*}$ as characterized in Proposition 2.2 into Proposition 2.1 and then the dual problem \ref{eqn:dual}.
\end{appendixproof}

\subsection{FBA}

\subsubsection{Inner Optimization Problem}

The robust FBA can be written as
\begin{equation*}\label{eqn:primal2}
 \sup_{Q \in \mathcal{U}_{\delta_2}(Q_N)} \mathbb{E}^Q [\langle Z^-, Y_{CF} \rangle]  \tag{P2}.
\end{equation*}

Now use recent duality results, noting the inner product $\langle \: ; \rangle$ satisfies the upper semicontinuous condition of the Lagrangian duality theorem, and cost function $c_S$ satisfies the non-negative lower semicontinuous condition (see \citet{blanchetFirst} Assumptions 1 \& 2, \citet{Gao16}). Hence the dual problem (to sup above) can be written as
\begin{equation*}\label{eqn:dual2}
  \inf_{\beta \geq 0} \: G(\beta) := \bigg[ \beta \delta_2 + \frac{1}{N} \sum_{i=1}^{N} \Psi_\beta(z^-_i,y^{cf}_i) \bigg]  \tag{D2}
\end{equation*}
where 
\begin{equation*}
\Psi_\beta(z^-_i,y^{cf}_i)  = \sup_{u \in \mathbb{R}^n_{-}, v \in {B^1_n}} [  \langle u,v \rangle - \beta c_{S_2}((u,v),(z^-_i,y^{cf}_i)) ] = \sup_{u \in \mathbb{R}^n_{-}, v \in {B^1_n}} [  \langle u, v \rangle - \beta( \langle u-z^-_i, u-z^-_i \rangle + S_2 \langle v-y^{cf}_i, v-y^{cf}_i \rangle ) ].
\end{equation*}


Now apply change of variables $w_1 = (u-z^-_i)$ and $w_2 = (v-y^{cf}_i)$ to get
\begin{equation*}
\Psi_\beta(z^-_i,y^{cf}_i)  = \sup_{w_1 \leq -z_i^-, w_2 \in {B^2_n}} [ \langle w_1+z^-_i, w_2+y^{cf}_i \rangle - \beta ( \langle w_1, w_1 \rangle + S_2 \langle w_2, w_2 \rangle ) ]
\end{equation*}
where sets $B^1_n$ and $B^2_n$ are defined as before. Following a similar approach as for FCA, it turns out that $\Psi_\beta$ can be expressed as original FBA plus the pointwise max of $(n+1)$ convex functions. The degenerate case $l=0$ is just a line of negative slope. The other $n$ cases are a (convex) piecewise line then hyperbola function plus a line of negative slope. $\Psi_\beta$ quantifies the adversarial move in FBA across both time and spatial dimensions while accounting for the cost via the $K$ terms.

\begin{proprep}
We have $\Psi_\beta(z^-_i,y^{cf}_i) = \, \langle z^-_i,y^{cf}_i \rangle + h_i(\beta)$ \\
where $h_i(\beta) = \big[ h_i(\beta,l^*) + \big( \sum_{k=1}^{l^*} z^-_{ik}  - \sum_{k=1}^{\|y^{cf}_i\|_1} z^-_{ik} \big) - \beta S_2 K \big] = \big[ \bigvee_{l=0}^n h_i(\beta,l) + \big( \sum_{k=1}^{l} z^-_{ik}  - \sum_{k=1}^{\|y^{cf}_i\|_1} z^-_{ik} \big) - \beta S_2 K \big]$.
Also $\|y^{cf}_i\|_1 \in \mathbb{Z^+}$, and  $K = | l - \| y^{cf}_i \|_1 | = \| w_2 \|_1 \geq 0, K \in \mathbb{Z^+}$.
Once $l^*$ is selected, $K := | l^* - \| y^{cf}_i \|_1 | = \| w_2^* \|_1$.
Continuing with the notational setup,
\begin{equation*}
h_i(\beta,l) = \sum_{k=1}^l g_{ik}(\beta) = \sum_{k=1}^l
\begin{cases}
-z_{ik}^- - \beta (z_{ik}^-)^2,	&  -z_{ik}^- \leq \frac{1}{2\beta} \\
\frac{1}{4\beta},					&  -z_{ik}^- > \frac{1}{2\beta} \, .
\end{cases}
\end{equation*}
Furthermore, $l^*$ is determined as
\begin{equation*}
l^* = \argmax_{l \in \{0,\dots,n\}} [ h_i(\beta,l) + \sum_{k=1}^l z^-_{ik} - \beta S_2 K ].
\end{equation*}
Recall $\bigvee_{l=0}^n h_i(\beta,l)$ denotes $\max_{l \in \{0,\dots,n\}} h_i(\beta,l)$.
\end{proprep}
\begin{proofsketch}
This result follows from jointly maximizing the adversarial funding exposure $w_1$ and the survival time index $w_2$.  The structure of $B^2_n$ allows us to decouple this joint maximization and find the critical point to maximize the quadratic in $w_1$ and write down the condition to select the optimal survival time index $l^*$. Finally, consider the two cases $w_2 = 0$ and $w_2 \neq 0$ and take the max to arrive at the solution. The constraint $w_1 \leq -z_i^-$ leads to a convex but piecewise structure for $h_i(\beta,l)$. The $K$ terms represent the cost associated with the worst case.
\end{proofsketch}
\begin{appendixproof}
\begingroup
\setlength{\parindent}{0pt}

The particular structure of $B^1_n$ and $B^2_n$ will be exploited to evaluate the $\sup$ above.
The analysis proceeds by considering different cases for optimal values $(w_1^{*}, w_2^{*})$.\\ 

$
\mathbf{Case \, 1}
$
\quad Suppose $w_2^{*} = 0 \implies  l = \| y^{cf}_i \|_1$. Then\\
\begin{equation*}
\Psi_\beta(z^-_i,y^{cf}_i)  = \, \langle z^-_i,y^{cf}_i \rangle + \sup_{w_1 \leq -z_i^-} [  \langle w_1, y^{cf}_i \rangle - \beta \langle w_1, w_1 \rangle ].
\end{equation*}
First look at the unconstrained problem,
\begin{equation*}
\tilde{\Psi}_\beta(z^-_i,y^{cf}_i)  = \, \langle z^-_i,y^{cf}_i \rangle + \sup_{w_1} [  \langle w_1, y^{cf}_i \rangle - \beta \langle w_1, w_1 \rangle ].
\end{equation*}

Applying the Cauchy-Schwarz Inequality gives
\begin{equation*}
\tilde{\Psi}_\beta(z^-_i,y^{cf}_i)  = \, \langle z^-_i,y^{cf}_i \rangle + \sup_{\| w_1 \|} [ \| w_1\| \| y^{cf}_i \| - \beta \| w_1 \|^2 ].
\end{equation*}
Evaluating the critical point $\|w_1^{*}\| = \frac{\|y^{cf}_i\|}{2\beta} \in \mathbb{R}_{+}$ for the quadratic gives
\begin{equation*}
\tilde{\Psi}_\beta(z^-_i,y^{cf}_i)  = \, \langle z^-_i,y^{cf}_i \rangle + \frac{ \| y^{cf}_i \|^2}{4\beta} = \langle z^-_i,y^{cf}_i \rangle + \frac{ \| y^{cf}_i \|_1}{4\beta}.
\end{equation*}
Now let us return to the constrained problem, $\Psi_\beta$.\\
Observing that only the first $l = \| y^{cf}_i \|_1$ components of $w_1 \text{ inside the sup are positive gives } \forall k \in \{1,\ldots,l\}$
\begin{equation*}
\sup_{w_1 \leq -z_i^- } [ \sum_{k=1}^{l} (w_{1k}) - \beta \langle w_1, w_1 \rangle  ]  = \sum_{k=1}^l \big[ \sup_{w_{1k} \leq -z^-_{ik}} w_{1k} - \beta ( w_{1k} )^2 \: \big].
\end{equation*}
Deduce that
\begin{equation*}
w_{1k}^* = \bigg[ -z_{ik}^- \wedge \frac{1}{2\beta} \bigg] \quad \forall k \in \{1,\ldots,l\}.
\end{equation*}
Therefore
\begin{equation*}
\Psi_\beta(z^-_i,y^{cf}_i)  = \, \langle z^-_i,y^{cf}_i \rangle + \sum_{k=1}^l \bigg[ -z_{ik}^- \wedge \frac{1}{2\beta} \bigg] - \beta \bigg[ -z_{ik}^- \wedge \frac{1}{2\beta} \bigg]^2.
\end{equation*}
Next, let us do some simplification for
\begin{equation*}
g_i(\beta) =  \sum_{k=1}^l \bigg[ -z_{ik}^- \wedge \frac{1}{2\beta} \bigg] - \beta \bigg[ -z_{ik}^- \wedge \frac{1}{2\beta} \bigg]^2.
\end{equation*}
Considering the two cases, it follows that:
\begin{equation*}
g_i(\beta) = \sum_{k=1}^l g_{ik}(\beta) = \sum_{k=1}^l
\begin{cases}
-z_{ik}^- - \beta (z_{ik}^-)^2,	&  -z_{ik}^- \leq \frac{1}{2\beta} \\
\frac{1}{4\beta},					&  -z_{ik}^- > \frac{1}{2\beta} \, .
\end{cases}
\end{equation*}

Note that $g_i(\beta)$ is a convex function! \\
In the degenerate case, $-z_{ik}^- = 0$, then $g_{ik}(\beta) = 0 \,\, \forall \beta \geq 0$, where $g_{ik}$ denotes the $k^{th}$ term in the sum. \\
Otherwise, $g_{ik}(\beta)$ is piecewise (line of negative slope for part 1, hyperbola for part 2) but still convex.
\begin{equation*}
g_{ik}'(\beta) = 
\begin{cases}
- (z_{ik}^-)^2,				&  -z_{ik}^- \leq \frac{1}{2\beta} \\
- \frac{1}{4\beta^2},		&  -z_{ik}^- > \frac{1}{2\beta} \, .
\end{cases}
\end{equation*}
Remarkably, these slopes are equal when $-z_{ik}^- = \frac{1}{2\beta}$ hence the convexity of $g_{ik}$ and thus $g_i$ holds.
Proceed to rewrite $\Psi_\beta$ as
\begin{equation*}
\Psi_\beta(z^-_i,y^{cf}_i)  = \, \langle z^-_i,y^{cf}_i \rangle + g_i(\beta).
\end{equation*}

\medskip
$
\mathbf{Case \, 2}
$
\quad Now consider $w_2^{*} \neq 0  \implies l \neq \| y^{cf}_i \|_1$.\\

Observe for $l = \|w_2 + y^{cf}_i \|_1 \geq 0$, \\
\begin{equation*}
\langle w_1+z^-_i, w_2 + y_i^{cf} \rangle = \sum_{k=1}^{l} (w_{1k} + z^-_{ik}).
\end{equation*}

The structure of finite set $B^2_n$ implies
\begin{equation*}
\Psi_\beta(z^-_i,y^{cf}_i) = \sup_{w_1 \in \mathbb{R}^n_{-}, l \in \{0,\ldots,n\}, l \neq \| y^{cf}_i \|_1} [ \sum_{k=1}^{l} (w_{1k} + z^-_{ik})  - \gamma ( \langle w_1, w_1 \rangle + S_2 K ) ].
\end{equation*}

Again, using that $B^2_n$ is a finite set, one can write
\begin{equation*}
\Psi_\beta(z^-_i,y^{cf}_i) = \max_{ l \in \{0,\ldots,n\}, l \neq \| y^{cf}_i \|_1} \sup_{w_1 \in \mathbb{R}^n_{-}} [ \sum_{k=1}^{l} (w_{1k} + z^-_{ik}) - \beta ( \langle w_1, w_1 \rangle + S_2 K ) ].
\end{equation*}
Observing that only the first $l$ components of $w_1 \text{ inside the sup are positive gives } \forall k \in \{1,\ldots,l\}$
\begin{equation*}
\sup_{w_1 \leq -z_i^- } [ \sum_{k=1}^{l} (w_{1k}) - \beta \langle w_1, w_1 \rangle  ]  = \sum_{k=1}^l \big[ \sup_{w_{1k} \leq -z^-_{ik}} w_{1k} - \beta ( w_{1k} )^2 \: \big].
\end{equation*}
Following the approach in \text{Case 1} above, define
\begin{equation*}
h_i(\beta,l) = \sum_{k=1}^l g_{ik}(\beta) = \sum_{k=1}^l
\begin{cases}
-z_{ik}^- - \beta (z_{ik}^-)^2,	&  -z_{ik}^- \leq \frac{1}{2\beta} \\
\frac{1}{4\beta},					&  -z_{ik}^- > \frac{1}{2\beta} \, .
\end{cases}
\end{equation*}
Furthermore, $l^*$ is determined as
\begin{equation*}
l^* = \argmax_{l \in \{0,\dots,n\}, l \neq \| y^{cf}_i \|_1} [ h_i(\beta,l) + \sum_{k=1}^l z^-_{ik} - \beta S_2 K ].
\end{equation*}
Proceed to write $\Psi_\beta$ as
\begin{equation*}
\Psi_\beta(z^-_i,y^{cf}_i) = [ h_i(\beta,l^*) + \sum_{k=1}^{l^*} z^-_{ik} - \beta S_2 K ].
\end{equation*}
This can be rewritten as
\begin{equation*}
\Psi_\beta(z^-_i,y^{cf}_i) =   \, \langle z^-_i,y^{cf}_i \rangle + \bigg[ h_i(\beta,l^*) + \bigg( \sum_{k=1}^{l^*} z^-_{ik}  - \sum_{k=1}^{\|y^{cf}_i\|_1} z^-_{ik} \bigg) - \beta S_2 K \bigg].
\end{equation*}
Introducing $h_i(\beta) := \big[ h_i(\beta,l^*) + \bigg( \sum_{k=1}^{l^*} z^-_{ik}  - \sum_{k=1}^{\|y^{cf}_i\|_1} z^-_{ik} \bigg) - \beta S_2 K \big]$,
\begin{equation*}
\Psi_\beta(z^-_i,y^{cf}_i) =   \, \langle z^-_i,y^{cf}_i \rangle + h_i(\beta).
\end{equation*}
$
\mathbf{Max}
$
\medskip

Taking the max values for $\Psi_\beta$ over cases $w_2^{*} = 0$ and $w_2^{*} \neq 0$ gives
\begin{equation*}
\Psi_\beta(z^-_i,y^{cf}_i) = \, \langle z^-_i,y^{cf}_i \rangle + g_i(\beta) \vee h_i(\beta).
\end{equation*}

Inspection suggests that $\Psi_\beta$ can be simplified further.
Observe that for $l^* = \| y^{cf}_i \|_1, \: h_i(\beta)$ evaluates to $g_i(\beta)$.
Let $l^*$ be determined as
\begin{equation*}
l^{*} = \argmax_{l \in \{0,\dots,n\}} [ h_i(\beta,l) + \sum_{k=1}^{l} (z^-_{ik}) - \beta S_2 K ]
\end{equation*}
and write
\begin{equation*}
\Psi_\beta(z^-_i,y^{cf}_i) = \, \langle z^-_i,y^{cf}_i \rangle + h_i(\beta).
\end{equation*}
Finally, note an alternate expression for $h_i(\beta)$ is $h_i(\beta) = \big[ \bigvee_{l=0}^n h_i(\beta,l) + \bigg( \sum_{k=1}^{l} z^-_{ik}  - \sum_{k=1}^{\|y^{cf}_i\|_1} z^-_{ik} \bigg) - \beta S_2 K \big] $.


%
\endgroup
\end{appendixproof}

\subsubsection{Outer Optimization Problem}

The goal now is to evaluate
\begin{equation*}
\inf_{\beta \geq 0} \: G(\beta) := \bigg[ \beta \delta_2 + \frac{1}{N} \sum_{i=1}^{N} \Psi_\beta(z^-_i,y^{cf}_i) \bigg]  
\end{equation*}
where 
\begin{equation*}
\Psi_\beta(z^-_i,y^{cf}_i)  = \, \langle z^-_i,y^{cf}_i \rangle + h_i(\beta) 
\end{equation*}
for $h_i(\beta) = \big[ h_i(\beta,l^*) +  \sum_{k=1}^{l^*} z^-_{ik} - \beta S_2 K \big] = \big[ \bigvee_{l=0}^n h_i(\beta,l) +  \sum_{k=1}^{l} z^-_{ik} - \beta S_2 K \big]$.\\

The convexity of the objective function $G(\beta)$ simplifies the task of solving this optimization problem. The first order optimality condition suffices. As $\Psi_\beta$ and hence $G(\beta)$ may have non-differentiable kinks due to the max functions, $\vee$, we characterize the optimality condition via subgradients. In particular, we look for $\beta^{*} \geq 0$ such that $0 \in \partial G(\beta^{*})$. Inspection of the asymptotic properties of $\Psi_\beta$ and its subgradients reveals that two cases are possible. Case 1 is $\partial G(\beta)$ consists of strictly positive elements hence $\beta^{*} = 0$. Case 2 is $\partial G(\beta)$will cross zero (as $\beta$ sweeps from $0$ to $\infty$) and hence $\beta^{*} \geq 0$.

\begin{proprep}
Let $\beta^{*} \in \left\{ \beta \geq 0 : 0 \in \partial G(\beta) \right\} \cup \{ \beta = 0 : 0 \notin \partial G(\beta)\} \,$ where \,
 $\partial \Psi_\beta = \partial h_i(\beta)$ \\
and $\partial G(\beta) = \delta_2 + \frac{1}{N} \sum_{i=1}^N \partial \Psi_\beta$;
$\partial h_i(\beta)= \mathbf{Conv} \cup \{ \partial h_i(\beta,l) - S_2 K \: | \:  h_i(\beta) = h_i(\beta,l) + \big( \sum_{k=1}^{l} z^-_{ik}  - \sum_{k=1}^{\|y^{cf}_i\|_1} z^-_{ik} \big) - \beta S_2 K ; \, l \in \{0,\dots,n\} \}$.
\end{proprep}

\begin{proofsketch}
This follows from application of standard properties of subgradients as well as inspection of the asymptotic properties of $\Psi_\beta$ and $\partial \Psi_\beta$. For Case 1, if $\partial G(\beta)$ consists of strictly positive elements then it is clear that $\beta^{*}$ attains the minimum. For Case 2, the asymptotic properties can be used to show that $\partial G(\beta)$ can't consist of strictly negative elements. For $\beta$ sufficiently large, for optimal $l^*$, either $l^* = 0 \implies 0 \in \partial \Psi_\beta$ or $l^* = \| y^{cf}_i \|_1 > 0 \implies \partial \Psi_\beta$ approaches $0 \implies \partial G(\beta)$ crosses zero.
\end{proofsketch}

\begin{appendixproof}
\begingroup
\setlength{\parindent}{0pt}

This follows from standard application of properties of convex functions and subgradients. 
First note that function $h_i(\beta)$ is convex in $\beta$ since it is the pointwise max of a finite set of convex functions plus a constant plus a negative linear term. 
So then $\Psi_\beta$ is also convex.
Using properties of subgradients, $\partial \Psi_\beta = \partial h_i(\beta)$;
$\partial h_i(\beta) = \mathbf{Conv} \cup \{ \partial h_i(\beta,l) - S_2 K \: | \:  h_i(\beta) = h_i(\beta,l) + \big( \sum_{k=1}^{l} z^-_{ik}  - \sum_{k=1}^{\|y^{cf}_i\|_1} z^-_{ik} \big) - \beta S_2 K ; \, l \in \{0,\dots,n\} \}$.
Continuing, $G(\beta)$ is convex since it is a linear term plus a sum of convex functions. It follows that $\partial G(\beta) = \delta_2 + \frac{1}{N} \sum_{i=1}^N \partial \Psi_\beta$.
Finally, we argue that $\beta^{*} \in \{ \beta \geq 0 : 0 \in \partial G(\beta) \} \cup \{ \beta = 0 : 0 \notin \partial G(\beta)\}$. 
Observe that for non-empty $\{ \beta \geq 0 : 0 \in \partial G(\beta) \}$, this defines $\beta^{*}$ due to convexity of $G(\beta)$. 
The claim is that for empty $\{ \beta \geq 0 : 0 \in \partial G(\beta) \}$ then $\beta^{*} = 0$. 
For the easier case, $0 < z_\beta \: \forall z_\beta \in \partial G(\beta), \: \forall \beta \: \geq 0$, it is clear that $\beta^{*} = 0$ minimizes $\inf_{\beta \geq 0} G(\beta)$. 
It remains to show that $0 > z_\beta \: \forall \, z_\beta \in \partial G(\beta), \: \forall \beta \: \geq 0$, does not occur. 
To elaborate, for $\beta > 0$ sufficiently large, $\|y^{cf}_i\|_1 > 0 \implies l^* = \| y^{cf}_i\|_1 \implies K = 0 \implies \exists \, z_\beta > -\delta_2$ such that $z_\beta \in \partial \Psi_\beta$.
To elaborate, for $\beta > 0$ sufficiently large, $\| y^{cf}_i \|_1 = 0 \implies l^* = 0 \implies K = 0, \Psi_\beta = 0, 0 = z_\beta > -\delta_2$ such that $z_\beta \in \partial \Psi_\beta$.
Hence we deduce $\exists \, z_\beta$ such that $0 < z_\beta + \delta_2 \in \partial G(\beta)$ and by continuity and convexity of $G(\beta)$ the claim holds. 

\endgroup
\end{appendixproof}

\begin{theoremrep}
The primal problem \ref{eqn:primal2} has solution 
$\big[ \beta^{*} \delta_2 + \frac{1}{N} \sum_{i=1}^{N} \Psi_{\beta^{*}}(z^-_i,y^{cf}_i) \big]$ \\
where $\beta^{*} \in \,  \left\{ \beta \geq 0 : 0 \in \partial G(\beta) \right\} \cup \{ \beta = 0 : 0 \notin \partial G(\beta)\}$ and
$\Psi_\beta(z^-_i,y^{cf}_i)  = \, \langle z^-_i,y^{cf}_i \rangle + h_i(\beta)$ where $h_i(\beta) = \big[ \bigvee_{l=0}^n h_i(\beta,l) + \big( \sum_{k=1}^{l} z^-_{ik}  - \sum_{k=1}^{\|y^{cf}_i\|_1} z^-_{ik} \big) - \beta S_2 K \big] $.
Expressed in terms of original FBA, this says
\begin{equation*}
\begin{split}
\sup_{Q \in \mathcal{U}_{\delta_2}(Q_N)} \mathbb{E}^Q [\langle Z^-,Y_{CF} \rangle] = \mathbb{E}^{Q_N} [\langle Z^-,Y_{CF} \rangle] + \beta^{*} \delta_2  \:  + 
\mathbb{E}^{Q_N} \big[ \bigvee_{l=0}^n h(\beta^{*},l) + \big( \sum_{k=1}^{l} Z^-_{k}  - \sum_{k=1}^{\|Y_{CF}\|_1} Z^-_{k} \big) - \beta^{*} S_2 K \big]
\end{split}
\end{equation*}
where the additional terms represent a penalty due to uncertainty in probability distribution, and 
\begin{equation*}
h(\beta,l) = \sum_{k=1}^l
\begin{cases}
-Z_{k}^- - \beta (Z_{k}^-)^2,	&  -Z_{k}^- \leq \frac{1}{2\beta} \\
\frac{1}{4\beta},					&  -Z_{k}^- > \frac{1}{2\beta} \, .
\end{cases}
\end{equation*}
\end{theoremrep}
\begin{proofsketch}
This result follows directly from the previous two propositions.
\end{proofsketch}
\begin{appendixproof}
This follows by direct substitution of $\beta^{*}$ as characterized in Proposition 2.4 into Proposition 2.3 and then the dual problem \ref{eqn:dual2}.
\end{appendixproof}

\subsection{FVA}

\subsubsection{Inner Optimization Problem}

The robust FVA can be written as
\begin{equation*}\label{eqn:primal3}
\sup_{\Phi \in \mathcal{U}_{\delta_3}(\Phi_N)} \mathbb{E}^\Phi [\langle Z, Y_{CF} \rangle]  \tag{P3}.
\end{equation*}

Similar to before, use recent duality results, noting the inner product $\langle \: ; \rangle$ satisfies the upper semicontinuous condition of the Lagrangian duality theorem, and cost function $c_S$ satisfies the non-negative lower semicontinuous condition (see \citet{blanchetFirst} Assumptions 1 \& 2, \citet{Gao16}). Hence the dual problem (to sup above) can be written as
\begin{equation*}\label{eqn:dual3}
 \inf_{\alpha \geq 0} \: F(\alpha) := \bigg[ \alpha \delta_3 + \frac{1}{N} \sum_{i=1}^{N} \Psi_\alpha(z_i,y^{cf}_i) \bigg]  \tag{D3}
\end{equation*}
where 
\begin{equation*}
\Psi_\alpha(z_i,y^{cf}_i)  = \sup_{u \in \mathbb{R}^n, v \in {B^1_n}} [  \langle u,v \rangle - \alpha c_{S_3}((u,v),(z_i,y^{cf}_i)) ] = \sup_{u \in \mathbb{R}^n, v \in {B^1_n}} [  \langle u, v \rangle - \alpha( \langle u-z_i, u-z_i \rangle + S_3 \langle v-y^{cf}_i, v-y^{cf}_i \rangle ) ].
\end{equation*}


Now apply change of variables $w_1 = (u-z_i)$ and $w_2 = (v-y^{cf}_i)$ to get
\begin{equation*}
\Psi_\alpha(z_i,y^{cf}_i)  = \sup_{w_1 \in \mathbb{R}^n, w_2 \in {B^2_n}} [ \langle w_1+z_i, w_2+y^{cf}_i \rangle - \alpha ( \langle w_1, w_1 \rangle + S_3 \langle w_2, w_2 \rangle ) ]
\end{equation*}
where the sets $B^1_n$ and $B^2_n$ are defined as before. It turns out that $\Psi_\alpha$ can be expressed as original FVA plus the pointwise max of $(n+1)$ convex functions. The degenerate case $l=0$ is just a line of negative slope. The other $n$ cases are a hyperbola plus a line of negative slope. $\Psi_\alpha$ quantifies the adversarial move in FVA across both time and spatial dimensions while accounting for the cost via the $K$ terms.


\begin{proprep}
We have \, $\Psi_\alpha(z_i,y^{cf}_i) = \, \langle z_i,y^{cf}_i \rangle + \big[  \frac{l^*}{4 \alpha} + \big(\sum_{k=1}^{l^*}z_{ik} - \sum_{k=1}^{\|y^{cf}_i\|_1} z_{ik}\big) - \alpha S_3 K \big]$ \\
where $l^{*} = \argmax_{l \geq 0} [ \frac{l}{4 \alpha} + \sum_{k=1}^{l} z_{ik} - \alpha S_3 K]$ and $l = \|w_2 + y^{cf}_i \|_1 \geq 0, \:  l \in \mathbb{Z^+}$.
Also $\|y^{cf}_i\|_1 \in \mathbb{Z^+}$, and  $K = | l - \| y^{cf}_i \|_1 | = \| w_2 \|_1 \geq 0, K \in \mathbb{Z^+}$.
Once $l^*$ is selected, $K := | l^* - \| y^{cf}_i \|_1 | = \| w_2^* \|_1$.
Alternatively, $\Psi_\alpha(z_i,y^{cf}_i) = \, \langle z_i,y^{cf}_i \rangle + \bigvee_{l=0}^n  h_\alpha(l)$ for \\ $h_\alpha(l) := \big[  \frac{l}{4 \alpha} + \big(\sum_{k=1}^{l}z_{ik} - \sum_{k=1}^{\|y^{cf}_i\|_1} z_{ik}\big) - \alpha S_3 K \big]$.
\end{proprep}
\begin{proofsketch}
This result follows from jointly maximizing the adversarial funding exposure $w_1$ and the survival time index $w_2$.  The structure of $B^2_n$ allows us to decouple this joint maximization and find the critical point to maximize the quadratic in $w_1$ and write down the condition to select the optimal survival time index $l^*$. Finally, consider the two cases $w_2 = 0$ and $w_2 \neq 0$ and take the max to arrive at the solution. The $K$ terms represent the cost associated with the worst case.
\end{proofsketch}
\begin{appendixproof}
\begingroup
\setlength{\parindent}{0pt}

The particular structure of $B^1_n$ and $B^2_n$ will be exploited to evaluate the $\sup$ above.
The analysis proceeds by considering different cases for optimal values $(w_1^{*}, w_2^{*})$.\\ \\
\medskip
$
\mathbf{Case \, 1}
$
\quad Suppose $w_2^{*} = 0 \implies  l = \| y^{cf}_i \|_1$. Then\\ 
\begin{equation*}
\Psi_\alpha(z_i,y^{cf}_i)  = \, \langle z_i,y^{cf}_i \rangle + \sup_{w_1 \in \mathbb{R}^n} [  \langle w_1, y^{cf}_i \rangle - \alpha \langle w_1, w_1 \rangle ].
\end{equation*}
Applying the Cauchy-Schwarz Inequality gives
\begin{equation*}
\Psi_\alpha(z_i,y^{cf}_i)  = \, \langle z_i,y^{cf}_i \rangle + \sup_{\| w_1 \|} [ \| w_1\| \| y^{cf}_i \| - \alpha \| w_1 \|^2 ].
\end{equation*}
Evaluating the critical point $\|w_1^{*}\| = \frac{\|y^{cf}_i\|}{2\alpha} \in \mathbb{R}_{+}$ for the quadratic gives
\begin{equation*}
\Psi_\alpha(z_i,y^{cf}_i)  = \, \langle z_i,y^{cf}_i \rangle + \frac{ \| y^{cf}_i \|^2}{4\alpha} = \langle z_i,y^{cf}_i \rangle + \frac{ \| y^{cf}_i \|_1}{4\alpha}.
\end{equation*} \\

$
\mathbf{Case \, 2}
$
\quad Now consider $w_2^{*} \neq 0 \implies l \neq \| y^{cf}_i \|_1 $. \\
Observe for $l = \|w_2 + y^{cf}_i \|_1 \geq 0$, \\
\begin{equation*}
\langle w_1+z_i, w_2 + y_i^{cf} \rangle = \sum_{k=1}^{l} (w_{1k} + z_{ik}).
\end{equation*}
The structure of finite set $B^2_n$ implies
\begin{equation*}
\Psi_\alpha(z_i,y^{cf}_i) = \sup_{w_1 \in \mathbb{R}^n, l \in \{0,\ldots,n\}, l \neq  \| y^{cf}_i \|_1} [ \sum_{k=1}^{l} (w_{1k} + z_{ik})  - \alpha ( \langle w_1, w_1 \rangle + S_3 K ) ].
\end{equation*}

Again, using that $B^2_n$ is a finite set, one can write
\begin{equation*}
\Psi_\alpha(z_i,y^{cf}_i) = \max_{ l \in \{0,\ldots,n\}, l \neq  \| y^{cf}_i \|_1} \sup_{w_1 \in \mathbb{R}^n} [ \sum_{k=1}^{l} (w_{1k} + z_{ik}) - \alpha ( \langle w_1, w_1 \rangle + S_3 K ) ].
\end{equation*}
Observing that only the first $l$ components of $w_1 \text{ inside the sup are positive gives } \forall k \in \{1,\ldots,l\}$
\begin{equation*}
\sup_{w_1 \in \mathbb{R}^n} [ \sum_{k=1}^{l} (w_{1k}) - \alpha \langle w_1, w_1 \rangle  ]  = l \times \sup_{w_{1k} \in \mathbb{R}} [ w_{1k} - \alpha ( w_{1k} )^2 ].
\end{equation*}
Evaluating at the critical point $w^{*}_{1k} = \frac{1}{2\alpha} \in \mathbb{R}_+$ for the above quadratic gives
\begin{equation*}
\sup_{w_{1k} \in \mathbb{R}} [ w_{1k} - \alpha (w_{1k}^2 ) ] = \frac{1}{4\alpha}.
\end{equation*}
Therefore one can write
\begin{equation*}
\Psi_\alpha(z_i,y^{cf}_i) = \max_{l \in \{0,\ldots,n\}, l \neq  \| y^{cf}_i \|_1} [ \frac{l}{4\alpha} + \sum_{k=1}^{l} (z_{ik}) - \alpha S_3 K ].
\end{equation*}

Furthermore, $l^{*}$ is determined as
\begin{equation*}
l^{*} = \argmax_{l \in \{0,\dots,n\}, l \neq  \| y^{cf}_i \|_1} [ \frac{l}{4\alpha} + \sum_{k=1}^{l} (z_{ik}) - \alpha S_3 K ].
\end{equation*}

Substituting back into expression for $\Psi_\alpha$ gives
\begin{equation*}
\Psi_\alpha(z_i,y^{cf}_i) = \, \langle z_i,y^{cf}_i \rangle + \bigg[  \frac{l^*}{4 \alpha} +  \bigg( \sum_{k=1}^{l^*} z_{ik}  - \sum_{k=1}^{\|y^{cf}_i\|_1} z_{ik} \bigg)  - \alpha S_3 K \bigg].
\end{equation*}

Finally, taking the max values for $\Psi_\alpha$ over cases $w_2^{*} = 0$ and $w_2^{*} \neq 0$ gives
\begin{equation*}
\Psi_\alpha(z_i,y^{cf}_i) = \, \langle z_i,y^{cf}_i \rangle + \bigg[\frac{ \| y^{cf}_i \|_1}{4 \alpha}\bigg] \vee  \bigg[  \frac{l^*}{4 \alpha} + \bigg( \sum_{k=1}^{l^*} z_{ik}  - \sum_{k=1}^{\|y^{cf}_i\|_1} z_{ik} \bigg) - \alpha S_3 K \bigg].
\end{equation*}
Observe that for $l^* = \| y^{cf}_i \|_1$, the last term in brackets $[ \, ; ]$ above evaluates to $\big[\frac{ \| y^{cf}_i \|_1}{4 \alpha}\big]$.
Let $l^*$ be determined as
\begin{equation*}
l^{*} = \argmax_{l \in \{0,\dots,n\}} [ \frac{l}{4\alpha} + \sum_{k=1}^{l} (z_{ik}) - \alpha S_3 K ]
\end{equation*}
and write
\begin{equation*}
\Psi_\alpha(z_i,y^{cf}_i) = \, \langle z_i,y^{cf}_i \rangle + \bigg[  \frac{l^*}{4 \alpha} + \bigg(\sum_{k=1}^{l^*}z_{ik} - \sum_{k=1}^{\|y^{cf}_i\|_1} z_{ik}\bigg) - \alpha S_3 K \bigg].
\end{equation*}

Alternatively, one can write
\begin{equation*}
\Psi_\alpha(z_i,y^{cf}_i) = \, \langle z_i,y^{cf}_i \rangle + \bigvee_{l=0}^n  \bigg[  \frac{l}{4 \alpha} + \bigg(\sum_{k=1}^{l}z_{ik} - \sum_{k=1}^{\|y^{cf}_i\|_1} z_{ik}\bigg) - \alpha S_3 K \bigg].
\end{equation*}

\endgroup
\end{appendixproof}

%
\subsubsection{Outer Optimization Problem}

The goal now is to evaluate
\begin{equation*}
 \inf_{\alpha \geq 0} \:  F(\alpha) := \bigg[ \alpha \delta_3 + \frac{1}{N} \sum_{i=1}^{N} \Psi_\alpha(z_i,y^{cf}_i) \bigg]  
\end{equation*}
where 
\begin{equation*}
\Psi_\alpha(z_i,y^{cf}_i)  = \, \langle z_i,y^{cf}_i \rangle + \bigvee_{l=0}^n h_\alpha(l) \:\: \text{for} \:\:  h_\alpha(l) := \big[  \frac{l}{4 \alpha} + \big(\sum_{k=1}^{l}z_{ik} - \sum_{k=1}^{\|y^{cf}_i\|_1} z_{ik}\big) - \alpha S_3 K \big].
\end{equation*}
The convexity of the objective function $F(\alpha)$ simplifies the task of solving this optimization problem. The first order optimality condition suffices. As $\Psi_\alpha$ and hence $F(\alpha)$ may have non-differentiable kinks due to the max functions, $\vee$, we characterize the optimality condition via subgradients. In particular, we look for $\alpha^{*} \geq 0$ such that $0 \in \partial F(\alpha^{*})$. Inspection of the asymptotic properties of $\Psi_\alpha$ and its subgradients reveals that $\partial F(\alpha)$ will cross zero (as $\alpha$ sweeps from $0$ to $\infty$) and hence $\alpha^{*} \geq 0$.


\begin{proprep}
Let $\alpha^{*} \in \,  \left\{ \alpha \geq 0: 0 \in \partial F(\alpha) \right\} \\ where \,\, \partial \Psi_\alpha = \mathbf{Conv} \cup \left\{ \partial h_\alpha(l) \: | \: \langle z_i,y^{cf}_i \rangle + h_\alpha(l) = \Psi_\alpha ; \, l \in \{0,\dots,n\} \right\}$
and $\partial F(\alpha) = \delta_3 + \frac{1}{N} \sum_{i=1}^N \partial \Psi_\alpha$\,.\\
\end{proprep}

\begin{proofsketch}
This follows from application of standard properties of subgradients as well as inspection of the asymptotic properties of $\Psi_\alpha$ and $\partial \Psi_\alpha$. For $\alpha$ sufficiently small, $\Psi_\alpha$ has a large positive value and $\partial \Psi_\alpha$ has a large negative derivative. For $\alpha$ sufficiently large, for optimal $l^*$, either $l^* = 0 \implies 0 \in \partial \Psi_\alpha$ or $l^* = \| y^{cf}_i \|_1 > 0 \implies \partial \Psi_\alpha$ approaches zero $\implies \partial F(\alpha)$ crosses zero.
\end{proofsketch}

\begin{appendixproof}
\begingroup
\setlength{\parindent}{0pt}

This follows from standard application of properties of convex functions and subgradients. 
First note that function $h_\alpha$ is convex in $\alpha$ since (for fixed $l$) it is the sum of a hyperbola plus a constant plus a negative linear term. 
So then $\Psi_\alpha$ is convex since it is the pointwise max of a finite set of convex functions plus a constant. 
Using properties of subgradients, one can write $\partial \Psi_\alpha = \mathbf{Conv} \cup \{ \partial h_\alpha(l) \: | \: \langle z_i,y^{cf}_i \rangle + h_\alpha(l) = \Psi_\alpha ; \, l \in \{0,\dots,n\} \}$. 
Furthermore $F(\alpha)$ is convex in $\alpha$ since it is a linear term plus a sum of convex functions, so one can write $\alpha^{*} \in \,  \left\{ \alpha : 0 \in \partial F(\alpha) \right\}$ and it follows that  $\partial F(\alpha) = \delta_3 + \frac{1}{N} \sum_{i=1}^N \partial \Psi_\alpha$.
Finally, we argue that $\alpha^{*} \geq 0$. 
For $\alpha > 0$ sufficiently small, $\exists \, z  < -\delta_3$ such that $z \in \partial \Psi_\alpha$ and for $\alpha > 0$ sufficiently large, $\exists \, z > -\delta_3$ such that $z \in \partial \Psi_\alpha$. 
To elaborate, for $\alpha > 0$ sufficiently large, $\| y^{cf}_i \|_1 > 0 \implies l^* = \| y^{cf} \|_1 \implies K = 0 \implies \exists \, z > -\delta_3$ such that $z \in \partial \Psi_\alpha$.
To elaborate, for $\alpha > 0$ sufficiently large, $\| y^{cf}_i \|_1 = 0 \implies l^* = 0 \implies K = 0, \Psi_\alpha = 0, 0 = z > -\delta_3$ such that $z \in \partial \Psi_\alpha$.
Hence we deduce $\partial F(\alpha)$ crosses the origin ( as $\alpha$ sweeps from $0$ to $\infty$ ). 
\endgroup
\end{appendixproof}


\begin{theoremrep}
The primal problem \ref{eqn:primal3} has solution 
$\big[ \alpha^{*} \delta_3 + \frac{1}{N} \sum_{i=1}^{N} \Psi_{\alpha^{*}}(z_i,y^{cf}_i) \big]$ \\
where $\alpha^{*} \in \,  \left\{ \alpha \geq 0: 0 \in \partial F(\alpha) \right\}$ and
$\Psi_{\alpha^{*}}(z_i,y^{cf}_i) = \, \langle z_i,y^{cf}_i \rangle + \bigvee_{l=0}^n h_{\alpha^{*}}(l) \:\: for \:\:  h_{\alpha^{*}}(l) := \big[  \frac{l}{4 \alpha^{*}} + \big(\sum_{k=1}^{l}z_{ik} - \sum_{k=1}^{\|y^{cf}_i\|_1} z_{ik}\big) - \alpha^{*} S_3 K \big]$. 
Expressed in terms of original FVA, this says
\begin{equation*}
\sup_{\Phi \in \mathcal{U}_{\delta_3}(\Phi_N)} \mathbb{E}^\Phi [\langle Z,Y_{CF} \rangle] = \mathbb{E}^{\Phi_N} [\langle Z,Y_{CF} \rangle] + \alpha^{*} \delta_3 +   \mathbb{E}^{\Phi_N} \big[ \bigvee_{l=0}^n \frac{l}{4 \alpha^{*}} + \big(\sum_{k=1}^{l}Z_k - \sum_{k=1}^{\|Y_{CF}\|_1} Z_k \big) - \alpha^{*} S_3 K  \big]
\end{equation*}
where the additional terms represent a penalty due to uncertainty in probability distribution.
\end{theoremrep}
\begin{proofsketch}
This follows directly from the previous two propositions.
\end{proofsketch}
\begin{appendixproof}
This follows by direct substitution of $\alpha^{*}$ as characterized in Proposition 2.6 into Proposition 2.5 and then the dual problem \ref{eqn:dual3}.
\end{appendixproof}

\endgroup

\section{Computational Study: Robust FVA and Wrong Way Funding Risk}
This computational study uses the Matlab Financial Instruments Toolbox and extends WWR portfolio analysis \citep[section 5.3]{brigo2013counterparty} to consider uncertainty in probability distribution. Other key concepts that will be leveraged in this section are the measure concentration results and association of Wasserstein radius $\delta$ with confidence level $1-\beta$ for some $\beta \in (0,1)$. As of July 10, 2019, 5y par interest rate swaps are 1.88\% (see \url{www.interestrateswapstoday.com}). The full table is shown below. Furthermore, Bloomberg shows U.S. CDX investment grade and high yield 5y credit default swap spreads as below.
\begin{table}[!h]
\begin{center}
\caption{Swap Rates}
\begin{tabular}{ |c|c|c|c|c|c|c|c| }
 \hline
Swap Tenor & 1y & 2y & 3y & 5y & 7y & 10y & 30y \\
 \hline
Swap Rate & 2.13\% & 1.95\% & 1.89\% & 1.88\% & 1.94\% & 2.05\% & 2.27\% \\
 \hline
\end{tabular}
\end{center}
\end{table}

\begin{table}[H]
\begin{center}
\caption{CDS Spreads}
\begin{tabular}{ |c|c|c| }
 \hline
CDX Index & IG & HY \\
 \hline
CDS Spread & 53 & 323 \\
 \hline
\end{tabular}
\end{center}
\end{table}

\noindent Referencing current (as of August 26, 2019) MarkIt funding spreads, the funding spread curves are set as below.
Unavailable quotes for high yield spreads are displayed as ``N/A".
\begin{table}[H]
\begin{center}
\caption{Funding Spreads}
\begin{tabular}{ |c|c|c|c|c|c|c| }
 \hline
Funding Tenor & 1y & 2y & 3y & 5y & 7y & 10y \\
 \hline
IG Spread & 0.13\% & 0.21\% & 0.31\% & 0.59\% & 0.84\% & 1.07\% \\
 \hline
HY Spread & N/A & N/A & 3.02\% & 3.62\% & 4.01\% & 4.09\% \\
 \hline
\end{tabular}
\end{center}
\end{table}

The computational studies in this section will investigate (and quantify) worst case FCA, FBA, and FVA for different market environments and portfolios of interest rate swaps. The current swaps curve (shown above) will be used in conjunction with monte carlo simulation of a one factor Hull-White model for interest rates. The funding spreads will be used in conjunction with a Libor Market Model (LMM) simulation of forward funding spreads. For this analysis, the same funding spreads will be used for both FCA and FBA calculations. The counterparty credit curve selection will vary between investment grade and high yield (as shown above). The different portfolio setups will be described in the following sections. All calculations are done in Matlab as an extension of the example provided in the financial instruments toolbox \citep {Matlab19}. The default Matlab settings for volatility and correlation parameters are used for the Hull-White and LMM term structure models. Independence between the funding cost, interest rate, and credit default factors is assumed for the joint simulation.

\subsection{Wasserstein Radius and Significance Levels}

As mentioned in Part 1 \citep{Singh19xva1}, a natural question to ask when computing worst case FVA is how to interpret the size of the Wasserstein radius $\delta$. Substantial research has been done and some key results are mentioned here. The following result is due to \citet{Fournier15} and the constants $c_1, c_2$ below can be calculated explicitly by following the proof:
\begin{equation*}
P[ D_c(\Phi,\Phi_N) \geq \delta)] \leq 
\begin{cases}
c_1 \exp(-c_2 N \delta^{\max\{n,2\}}) & \text{if} \: \delta \leq 1, \\
c_1 \exp(-c_2 N \delta^a) & \text{if} \: \delta > 1
\end{cases}
\end{equation*}
$\forall N \geq 1, n \neq 2, \, \text{and} \, \delta > 0$ where $c_1 > 0, c_2 > 0$ depend only on $a$, $A$, and $n$.\\
\citet{Esfahani17} discuss how equating the RHS above to $\beta$ and solving for $\delta$ gives
\begin{equation*}
\delta_N(\beta) =
\begin{cases}
(\frac{\log(c_1\beta^{-1}}{c_2N})^{1/\max\{n,2\}} & \text{if} \: N \geq \frac{\log(c_1 \beta^{-1})}{c_2}, \\
(\frac{\log(c_1\beta^{-1}}{c_2N})^{1/a} & \text{if} \: N < \frac{\log(c_1 \beta^{-1})}{c_2} 
\end{cases}
\end{equation*}
however these bounds are overly conservative, and result in a radius $\delta^*$ much larger than necessary.\par

As an alternative approach, we follow a method that provides a more explicit mapping between $\delta$ and $\beta$ \citep[section 3]{Carlsson2018}. Theorem 6.15 of \citet{Villani08} gives a bound on Wasserstein distance between two pdfs $\Phi, \Phi'$ as
\begin{equation*}
D(\Phi,\Phi') \leq \iint\limits_\mathcal{R} \| x_0 - x \| \cdot | \Phi(x) - \Phi'(x) | dA.
\end{equation*}
\citet{Carlsson2018} get the result
\begin{equation*}
P(D(\Phi,\Phi_N) \geq \delta)  \lesssim \exp \left( -N \frac{8r -2 \sqrt{16r^2 + 16r\delta + 24r + 12\delta+9} +4\delta+6 }{3+4r } \right)
\end{equation*}
where
\begin{equation*}
r = \max_{x_0 \in \mathcal{R},x \in \mathcal{R}} \|x - x_0\|
\end{equation*}
denotes the radius of domain $\mathcal{R}$.
This is the characterization which is used in this study.
Therefore, for a desired significance (confidence) level $\beta \in (0,1)$, find $\delta_\beta$ such that 
\begin{equation*}
1 - \beta = \exp \left( -N \frac{8r -2 \sqrt{16r^2 + 16r\delta_\beta + 24r + 12\delta_\beta+9} +4\delta_\beta+6 }{3+4r } \right).
\end{equation*}

In our problem setting we use $r_N$ (computed using the empirical domain $\mathcal{R}_N$) as the discrete approximation to $r$, which is difficult to bound. \par

\subsection{FCA}

\subsubsection{Portfolio of Interest Rate Swaps, Investment Grade Counterparty and Firm}
The portfolio here consists of a dozen interest rate swaps, with a mix of receving fixed and paying fixed swaps, at different coupons, maturities, and notionals. The fixed coupons range between 2\% and 2.5\%, the maturities range between 4y and 12y, the notionals range between 400k USD and 1mm USD. The investment grade counterparty and firm credit spreads are set to 50 basis points. The table of confidence levels $\beta$ and their corresponding Wasserstein radii $\delta$ follows. 
\begin{table}[H]
\begin{center}
\caption{FCA Investment Grade Wasserstein Radii}
\begin{tabular}{ |c|c|c|c|c|c|c| }
 \hline
Confidence Level & 0.80 & 0.85 & 0.90 & 0.95 & 0.99 & 0.999 \\
 \hline
W Radius delta & 1.2 & 1.3 & 1.5 & 1.7 & 2.1 & 2.6 \\
 \hline
\end{tabular}
\end{center}
\end{table}

The scale factor $S_1$ is set (by default) to 1 and the portfolio exposures are scaled to be in units of thousands of dollars. 
Again, the intent of scaling is to provide appropriate penalty to the adversarial change in joint distribution of portfolio funding exposures and default times that promotes worst case FCA and wrong way risk. Further work may conduct a sensitivity analysis regarding the pairings of $S_1$ and units of portfolio exposures to investigate suitable (unsuitable) ranges that preserve (distort) the shape of the robust FCA profile.
Matlab plots characterizing the FCA exposure profile and trajectory of worst case FCA as a function of Wasserstein radius are shown. Again, we think about worst case FCA (which incorporates joint survival probability) as compared to the funding PFE (potential future exposure) which shows tail percentiles of funding exposure (not scaled by joint survival probability). \par

\begin{figure}[!htb]
\caption{Swaps Portfolio Positive Exposure Profiles}
\centerline{\scalebox{0.4}[0.2]{\includegraphics{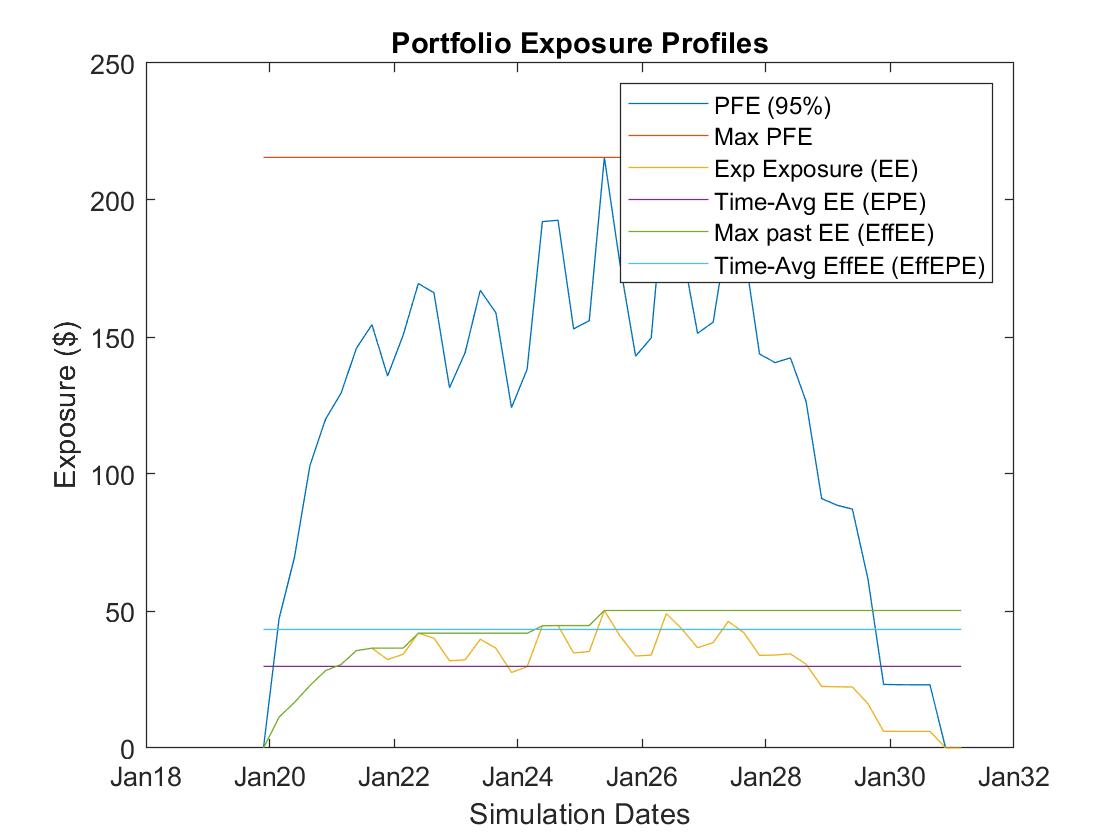}}}
\end{figure}

\begin{figure}[H]
\caption{Swaps Portfolio IG FCA Exposure Profiles}
\centerline{\scalebox{0.4}[0.2]{\includegraphics{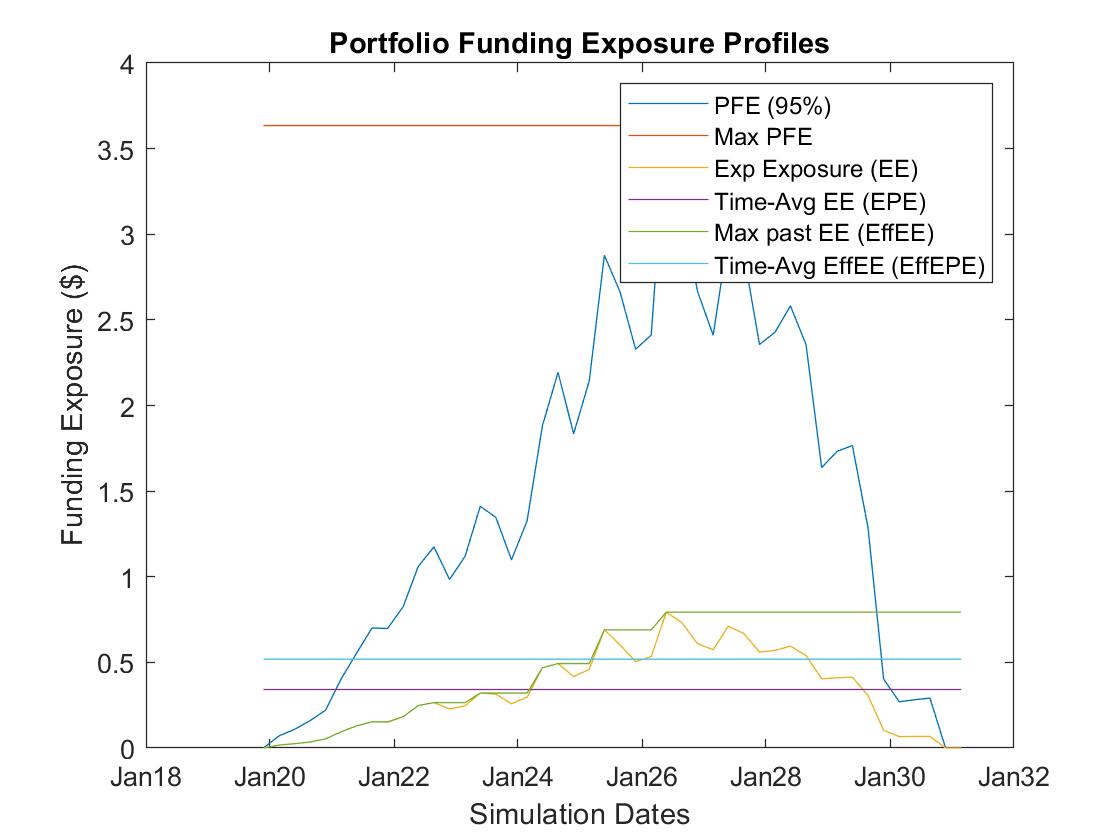}}}
\end{figure}

\begin{figure}[!htb]
\caption{Swaps Portfolio Worst Case IG FCA Profile}
\centerline{\scalebox{0.4}[0.2]{\includegraphics{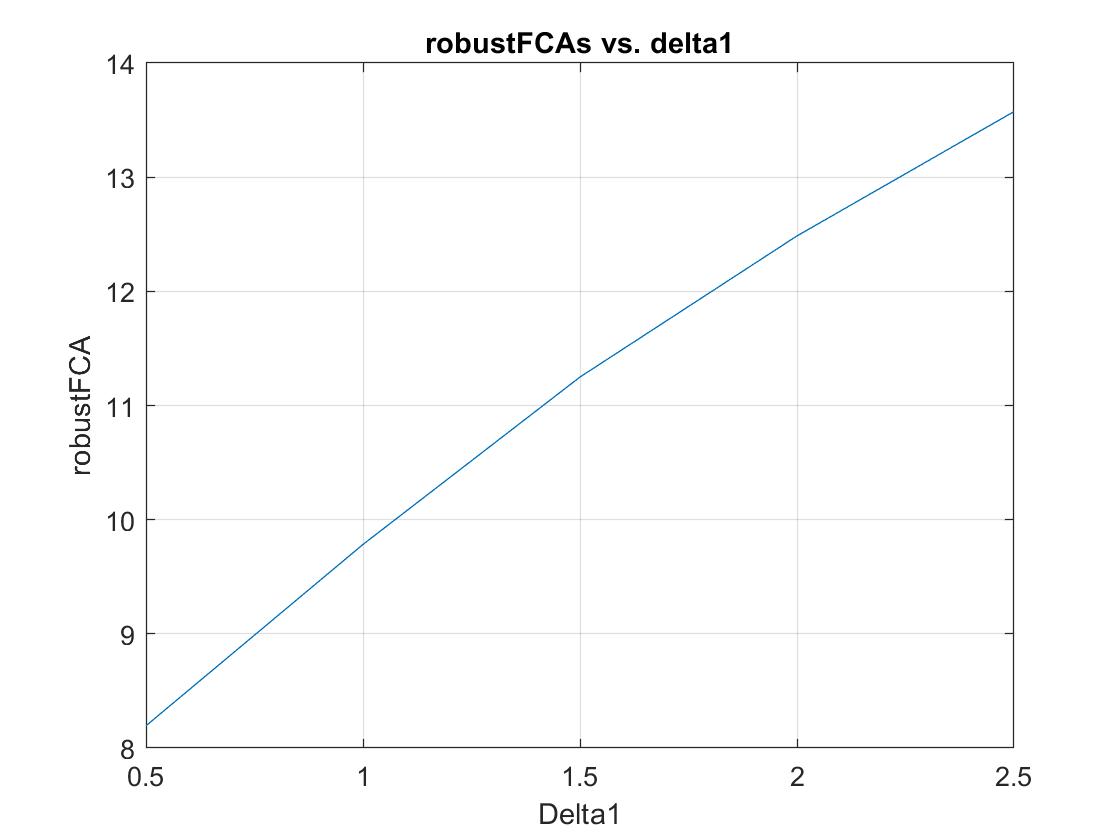}}}
\end{figure}


The baseline FCA for this portfolio is small (about 2.2k USD) and represents the dot product of the discounted portfolio funding exposure profile times joint survival probability. The worst case FCA curve is shown below. Note the worst case FCA is approximately 70\% the size of integrated (lifetime) FCA PFE for Wasserstein radius $\delta$ about 1.7 which maps to a significance level around $95\%$. So the takeaway here is worst case FCA is still a significant percentage of funding PFE for swap portfolios with low counterparty default curves (investment grade).\par



\subsubsection{Portfolio of Interest Rate Swaps, High Yield Counterparty and Firm}
The reference portfolio here consists of a dozen interest rate swaps, with a mix of receving fixed and paying fixed swaps, at different coupons, maturities, and notionals. The fixed coupons range between 2\% and 2.5\%, the maturities range between 4y and 12y, the notionals range between 400k USD and 1mm USD. The high yield counterparty and firm credit spreads are set to 320 basis points. The table of confidence levels $\beta$ and their corresponding Wasserstein radii $\delta$ is shown below. For the same reference portolio, and same set of monte carlo interest rate paths, the max interest rate exposures are the same. However, the high yield credit spreads and funding costs expand the Wasserstein radii. \par
\begin{table}[h]
\begin{center}
\caption{FCA High Yield Wasserstein Radii}
\begin{tabular}{ |c|c|c|c|c|c|c| }
 \hline
Confidence Level & 0.80 & 0.85 & 0.90 & 0.95 & 0.99 & 0.999 \\
 \hline
W Radius delta & 3.1 & 3.4 & 3.7 & 4.2 & 5.3 & 6.5 \\
 \hline
\end{tabular}
\end{center}
\end{table}


\begin{figure}[!htb]
\caption{Swaps Portfolio Positive Exposure Profiles}
\centerline{\scalebox{0.4}[0.21]{\includegraphics{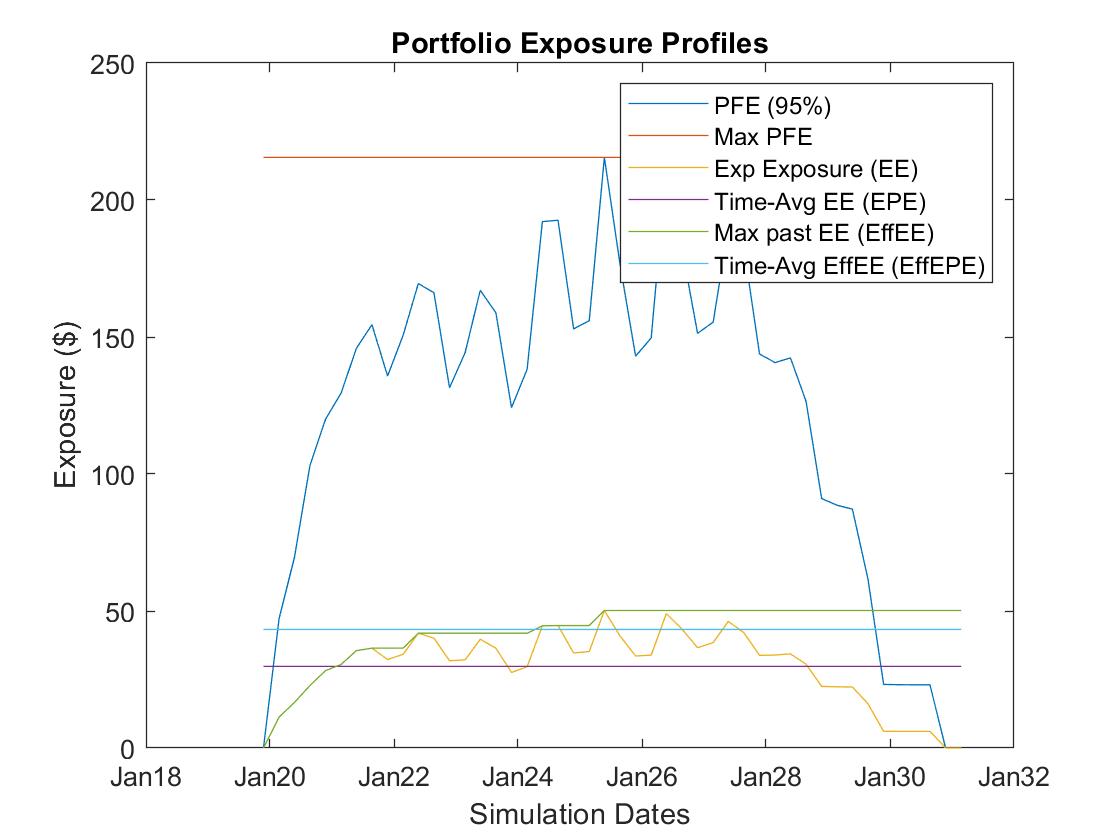}}}
\end{figure}

\begin{figure}[!htb]
\caption{Swaps Portfolio HY FCA Exposure Profiles}
\centerline{\scalebox{0.4}[0.21]{\includegraphics{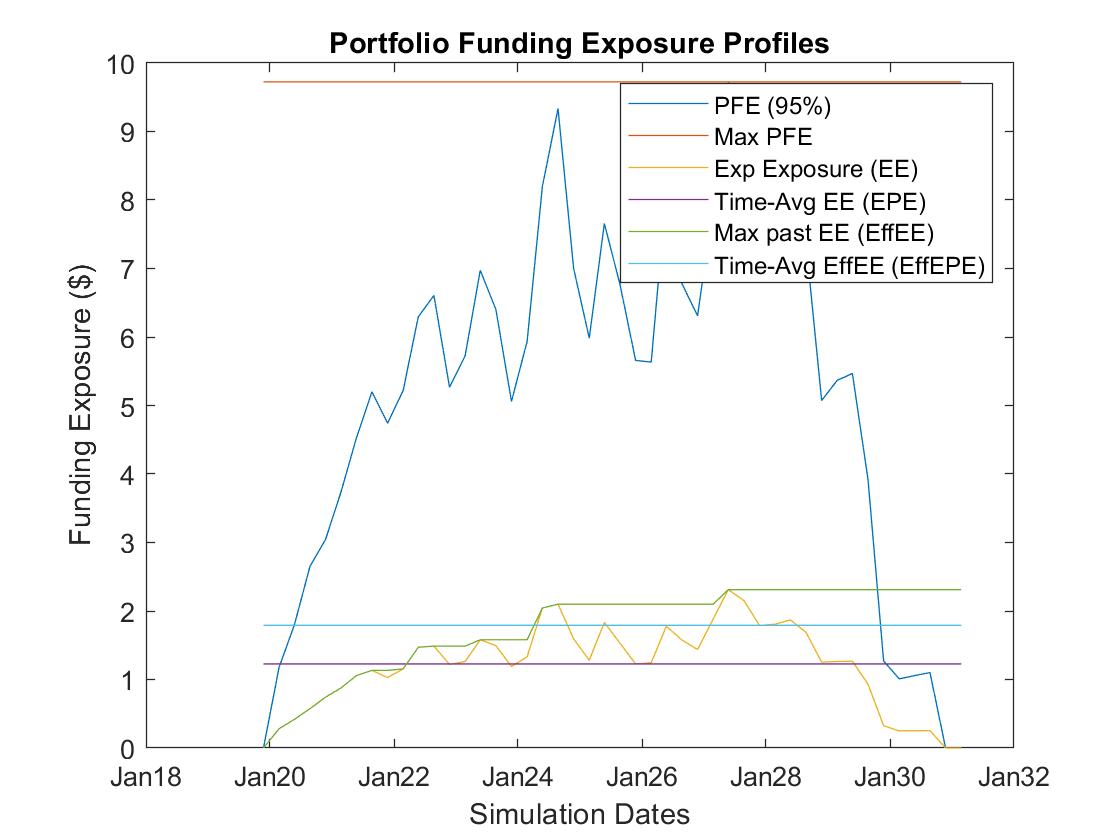}}}
\end{figure}

\begin{figure}[H]
\caption{Swaps Portfolio Worst Case HY FCA Profile}
\centerline{\scalebox{0.4}[0.2]{\includegraphics{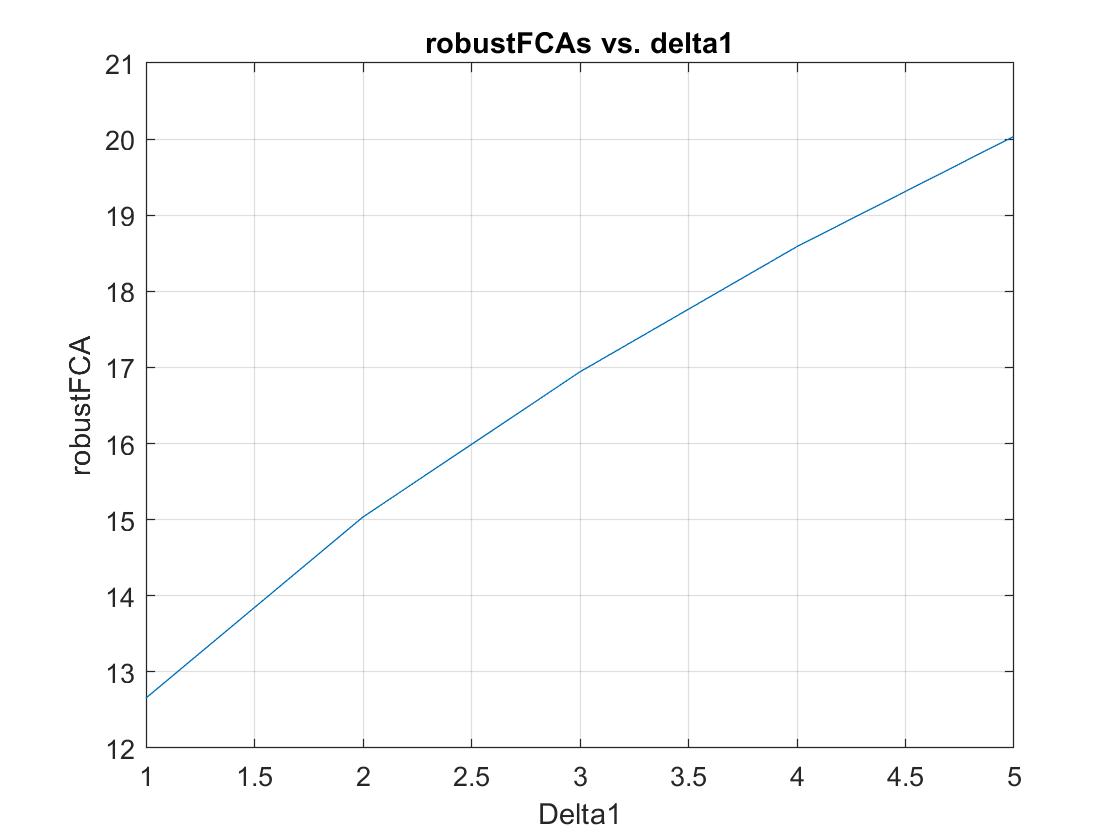}}}
\end{figure}



The baseline FCA for this portfolio is higher, around 65k USD and represents the dot product of the discounted portfolio funding exposure profile times joint survival probability. The worst case FCA curve is shown. Note the worst case FCA is still approximately 33\% the size of integrated (lifetime) FCA PFE (Potential Future Exposure) for Wasserstein radius $\delta$ about 4.2 which maps to a significance level around $95\%$. So the takeaway here is worst case FCA is a significant percentage of funding PFE for swap portfolios with moderately high counterparty and firm default curves. \par

\subsection{FBA}

\subsubsection{Portfolio of Interest Rate Swaps, Investment Grade Counterparty and Firm}

The portfolio here consists of a dozen interest rate swaps, with a mix of receving fixed and paying fixed swaps, at different coupons, maturities, and notionals. The fixed coupons range between 2\% and 2.5\%, the maturities range between 4y and 12y, the notionals range between 4mm USD and 10mm USD. The investment grade counterparty and firm credit spreads are set to 50 basis points. The table of confidence levels $\beta$ and their corresponding Wasserstein radii $\delta$ follows. \par
\begin{table}[h]
\begin{center}
\caption{FBA Investment Grade Wasserstein Radii}
\begin{tabular}{ |c|c|c|c|c|c|c| }
 \hline
Confidence Level & 0.80 & 0.85 & 0.90 & 0.95 & 0.99 & 0.999 \\
 \hline
W Radius delta & 1.0 & 1.1 & 1.2 & 1.3 & 1.7 & 2.1 \\
 \hline
\end{tabular}
\end{center}
\end{table}

The scale factor $S_2$ is set (by default) to 1 and the portfolio exposures are scaled to be in units of thousands of dollars. Same comments as above, regarding scaling, apply.
Matlab plots characterizing the FBA exposure profile and trajectory of worst case FBA as a function of Wasserstein radius are shown.  \par

\begin{figure}[!htb]
\caption{Swaps  Portfolio Negative Exposure Profiles}
\centerline{\scalebox{0.4}[0.2]{\includegraphics{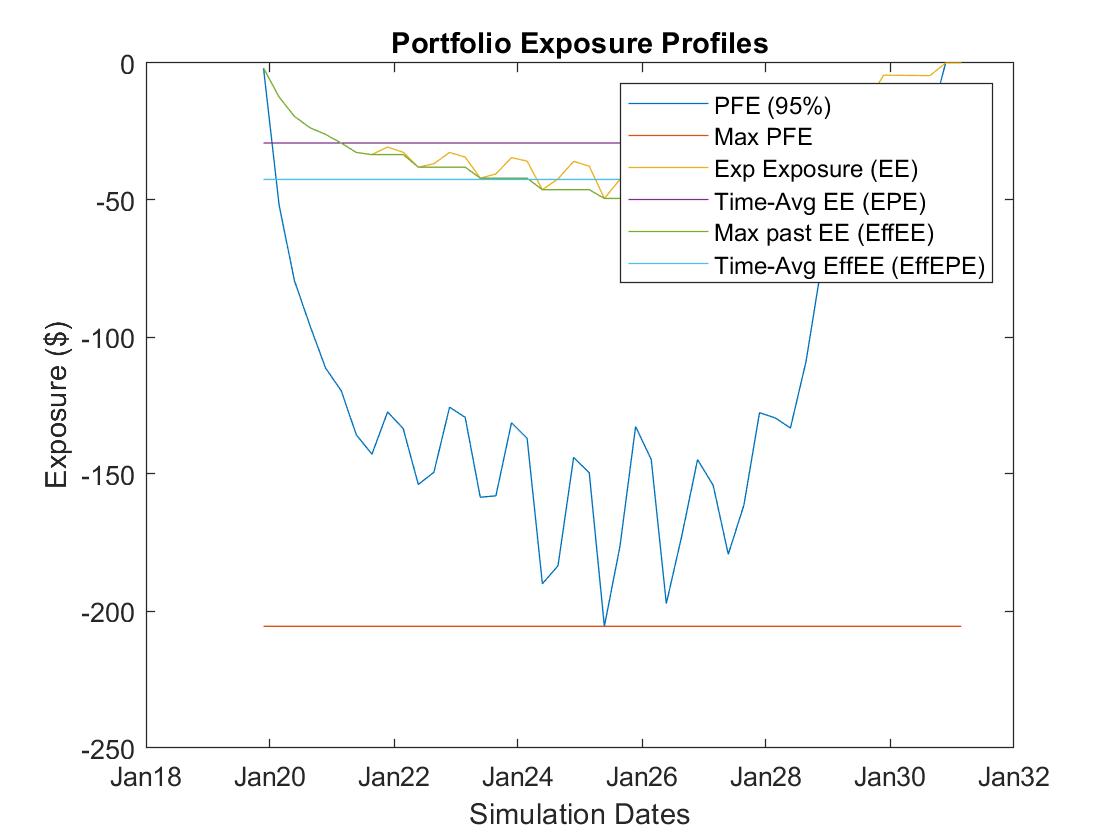}}}
\end{figure}

\begin{figure}[!htb]
\caption{Swaps Portfolio IG FBA Exposure Profiles}
\centerline{\scalebox{0.4}[0.2]{\includegraphics{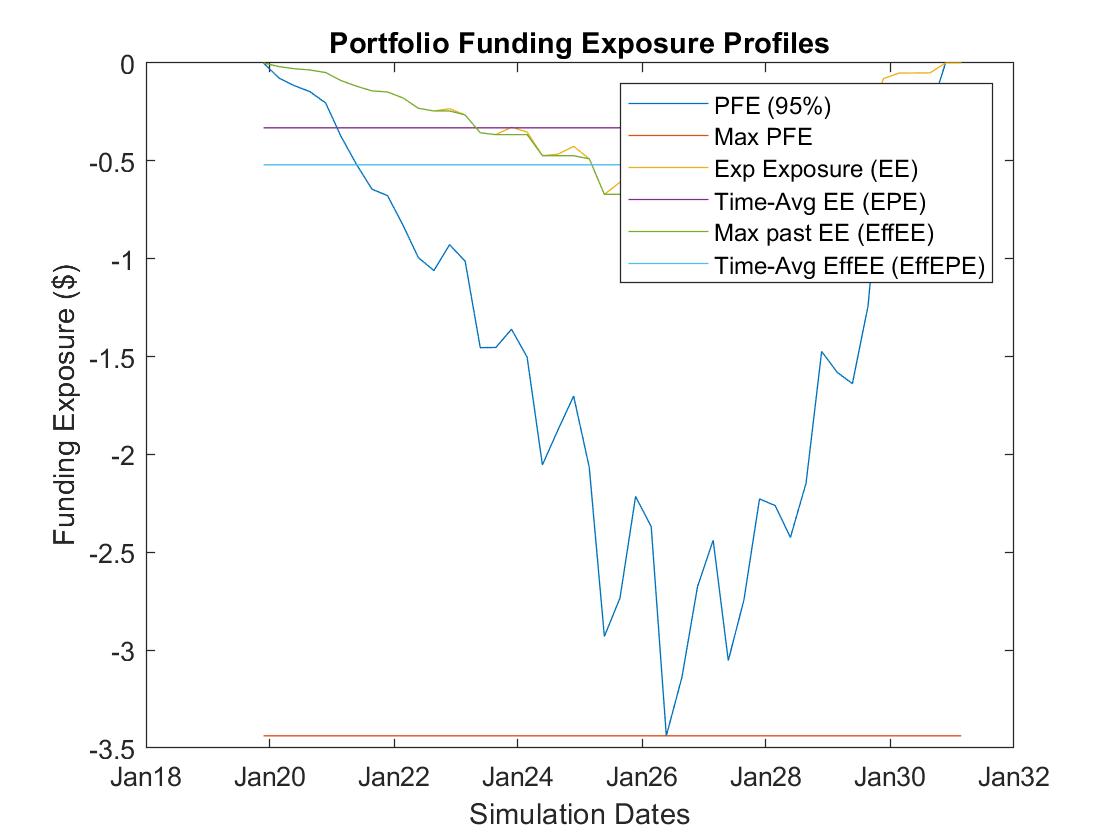}}}
\end{figure}

\begin{figure}[H]
\caption{Swaps Portfolio Worst Case IG FBA Profile}
\centerline{\scalebox{0.4}[0.2]{\includegraphics{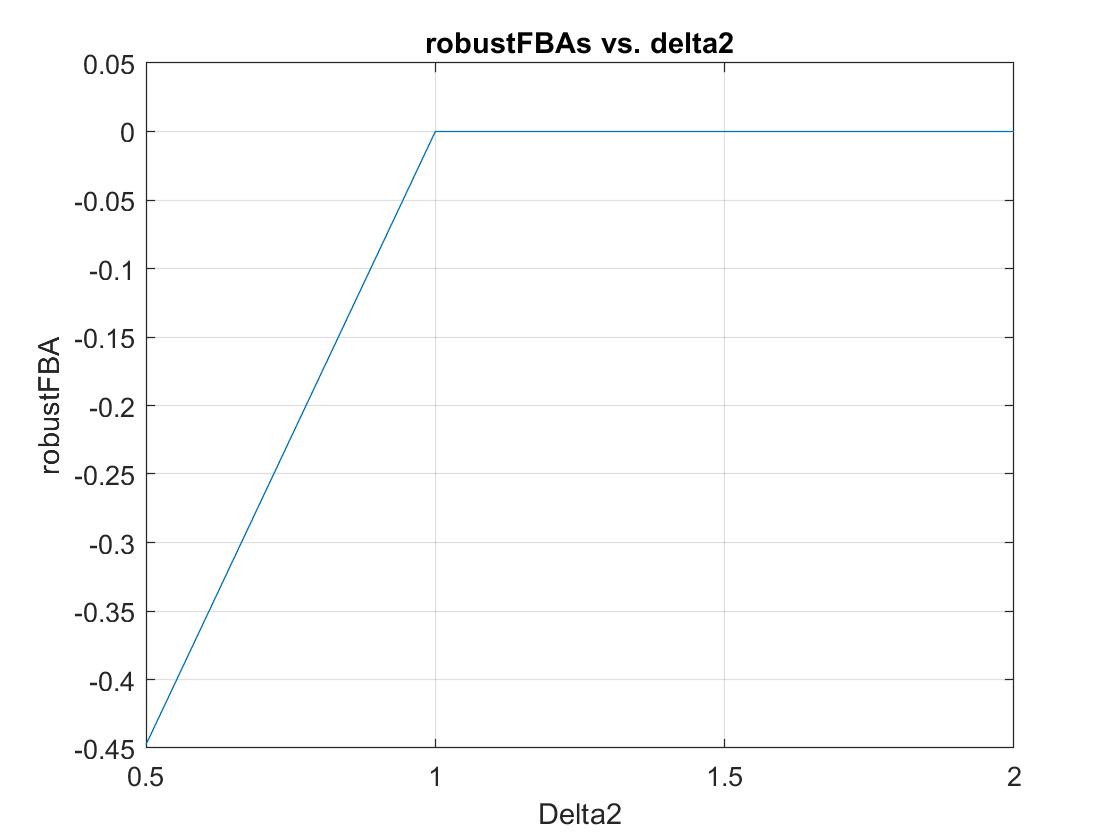}}}
\end{figure}


The baseline FBA for this portfolio is -2k USD and represents the dot product of the discounted negative funding portfolio exposure profile times joint survival probability. The worst case FBA plot is shown. The plot illustrates that worst case FBA quickly attains its lower bound (in magnitude) of zero (no funding benefit to the firm for FBA).

%
%
%


\subsubsection{Portfolio of Interest Rate Swaps, High Yield Counterparty and Firm}

The reference portfolio here is the same one used in the previous subsection, albeit with notionals from 4mm to 10mm USD. The high yield counterparty and firm credit spreads are set to 320 basis points. The table of confidence levels $\beta$ and their corresponding Wasserstein radii $\delta$ is shown.  For the same reference portolio, and same set of monte carlo interest rate paths, the max interest rate exposures are the same. However, the high yield credit spreads and funding costs expand the Wasserstein radii. \par
\begin{table}[h]
\begin{center}
\caption{FBA High Yield Wasserstein Radii}
\begin{tabular}{ |c|c|c|c|c|c|c| }
 \hline
Confidence Level & 0.80 & 0.85 & 0.90 & 0.95 & 0.99 & 0.999 \\
 \hline
W Radius delta & 2.0 & 2.2 & 2.4 & 2.8 & 3.4 & 4.3 \\
 \hline
\end{tabular}
\end{center}
\end{table}
A series of matlab plots characterizing the FBA exposure profile and trajectory of worst case FBA as a function of Wasserstein radius is shown.  

\begin{figure}[!htb]
\caption{Swaps Portfolio Negative Exposure Profiles}
\centerline{\scalebox{0.4}[0.2]{\includegraphics{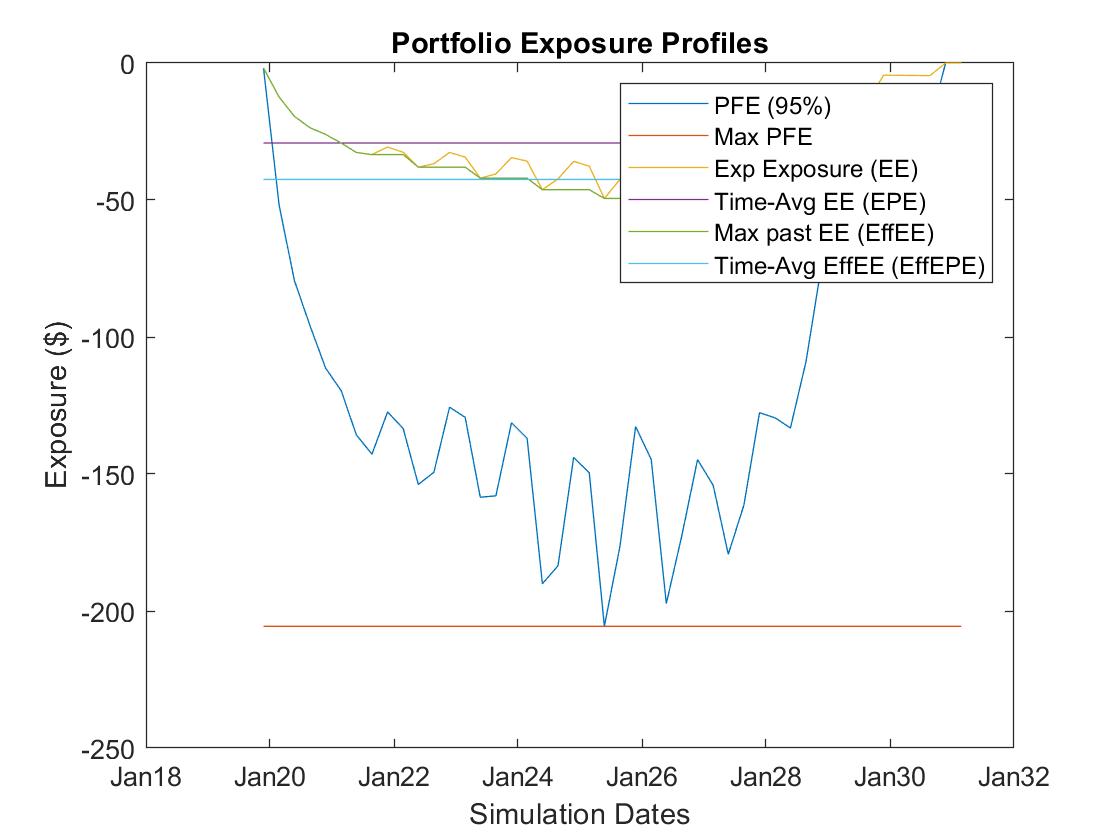}}}
\end{figure}

\begin{figure}[H]
\caption{Swaps Portfolio HY FBA Exposure Profiles}
\centerline{\scalebox{0.4}[0.2]{\includegraphics{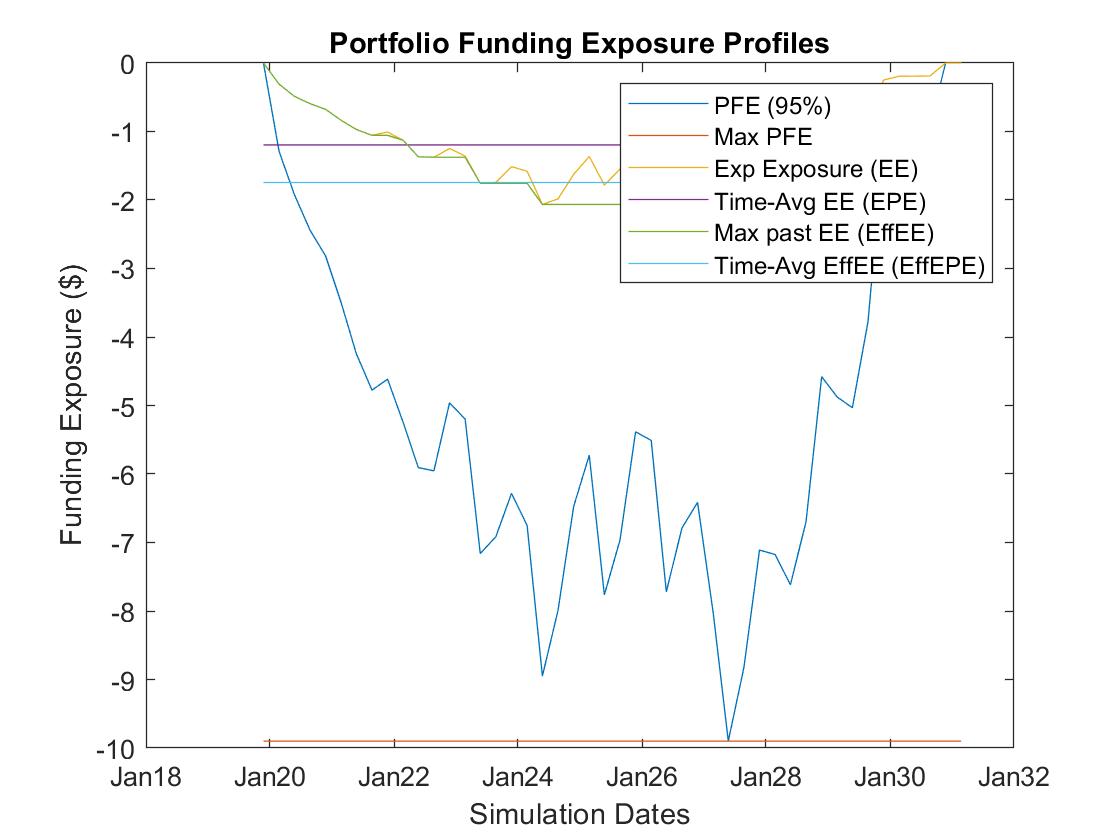}}}
\end{figure}

\begin{figure}[H]
\caption{Swaps Portfolio Worst Case HY FBA Profile}
\centerline{\scalebox{0.4}[0.2]{\includegraphics{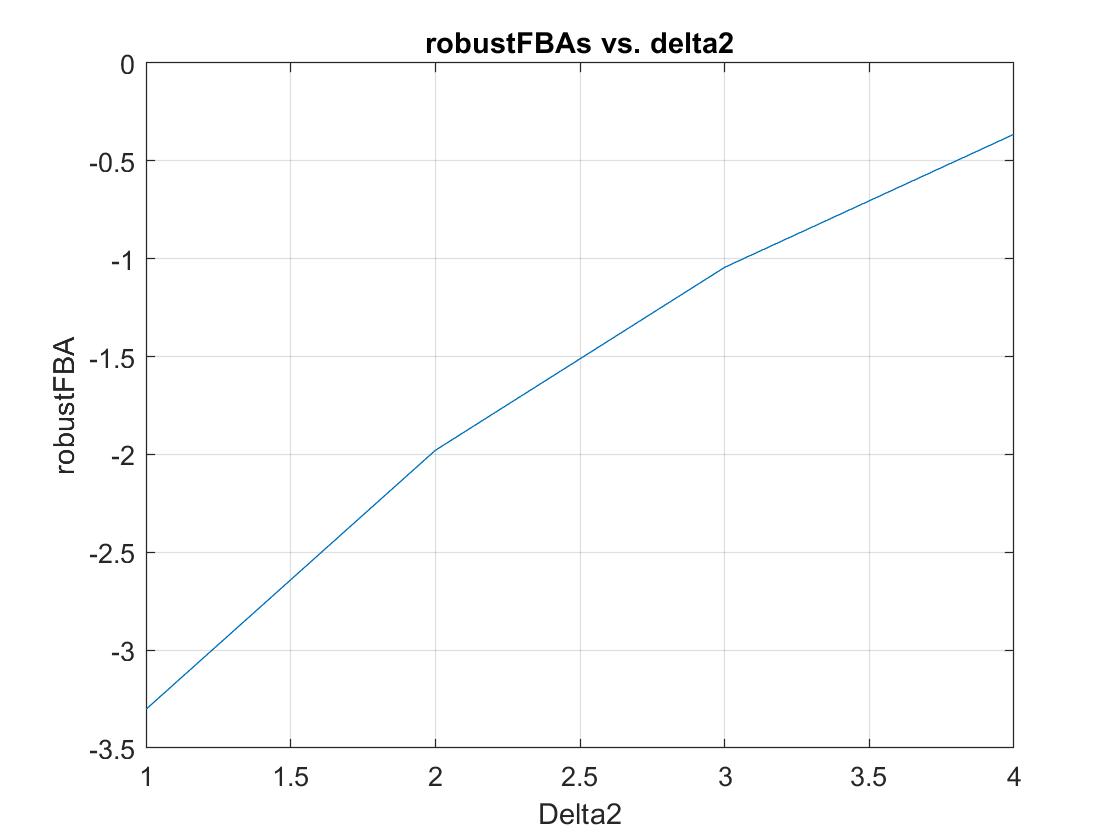}}}
\end{figure}


The baseline FBA for this portfolio is -6k USD and represents the dot product of the discounted negative portfolio funding exposure profile times joint survival probability. The worst case FBA plot is shown. Once again, the plot illustrates that worst case FBA moves towards its lower bound (in magnitude) of zero (no funding benefit to the firm for FBA). Note the worst case FBA is just 2.1\% the size of integrated (lifetime) FBA PFE (Potential Future Exposure) for Wasserstein radius $\delta$ about 2.8 which maps to a significance level around $95\%$. So it accelerates towards the lower bound (in magnitude) of zero.



\subsection{FVA}

\subsubsection{Portfolio of Interest Rate Swaps, Investment Grade Counterparty and Firm}

The corresponding FCA portfolio is used for comparison. The portfolio consists of a dozen interest rate swaps, with a mix of receving fixed and paying fixed swaps, at different coupons, maturities, and notionals. The fixed coupons range between 2\% and 2.5\%, the maturities range between 4y and 12y, the notionals range between 400k USD and 1mm USD. The investment grade counterparty and firm credit spreads are set to 50 basis points. The table of confidence levels $\beta$ and their corresponding Wasserstein radii $\delta$ follows. \par
\begin{table}[h]
\begin{center}
\caption{FVA Investment Grade Wasserstein Radii}
\begin{tabular}{ |c|c|c|c|c|c|c| }
 \hline
Confidence Level & 0.80 & 0.85 & 0.90 & 0.95 & 0.99 & 0.999 \\
 \hline
W Radius delta & 1.4 & 1.5 & 1.7 & 1.9 & 2.4 & 2.9 \\
 \hline
\end{tabular}
\end{center}
\end{table}

The scale factor $S_3$ is set (by default) to 1 and the portfolio exposures are scaled to be in units of thousands of dollars. Same comments as above, for FCA and FBA, regarding scaling, apply. Matlab plots characterizing the FVA positive and negative exposure profiles and trajectory of worst case FVA as a function of Wasserstein radius are shown.  \par


\begin{figure}[!htb]
\caption{Swaps Portfolio Positive Exposure Profiles}
\centerline{\scalebox{0.4}[0.2]{\includegraphics{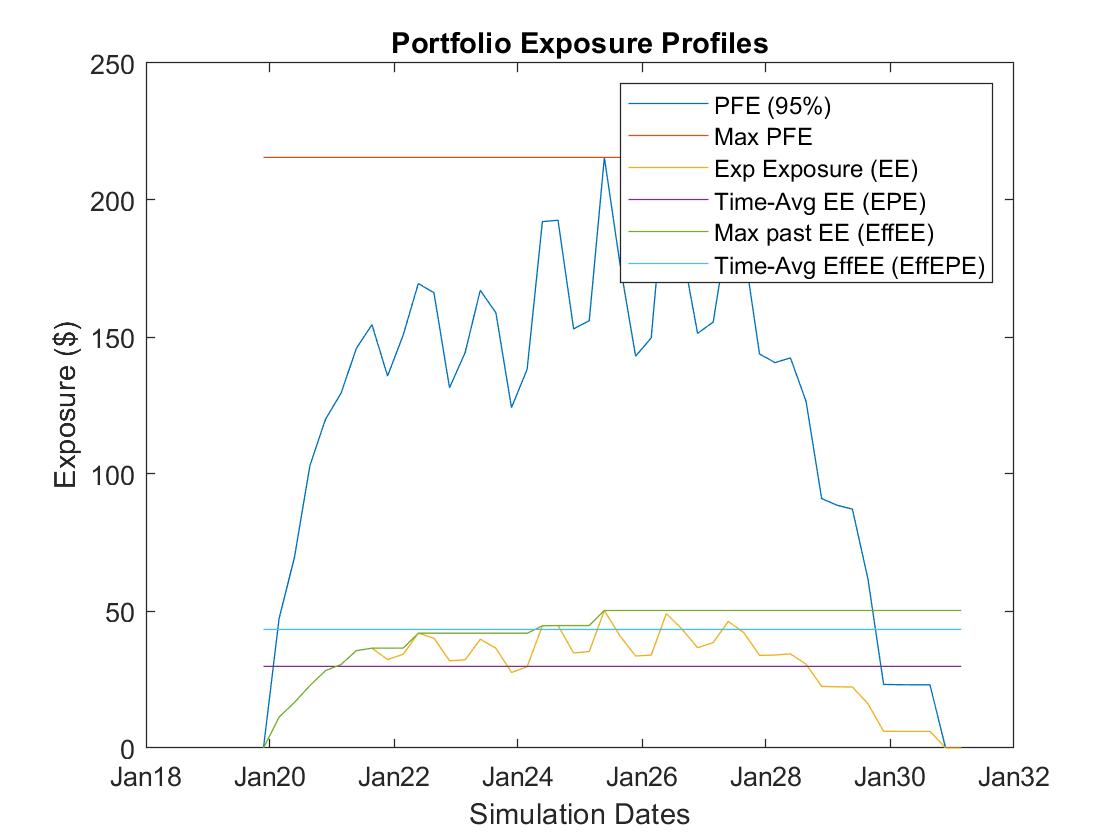}}}
\end{figure}

\begin{figure}[!htb]
\caption{Swaps Portfolio Negative Exposure Profiles}
\centerline{\scalebox{0.4}[0.2]{\includegraphics{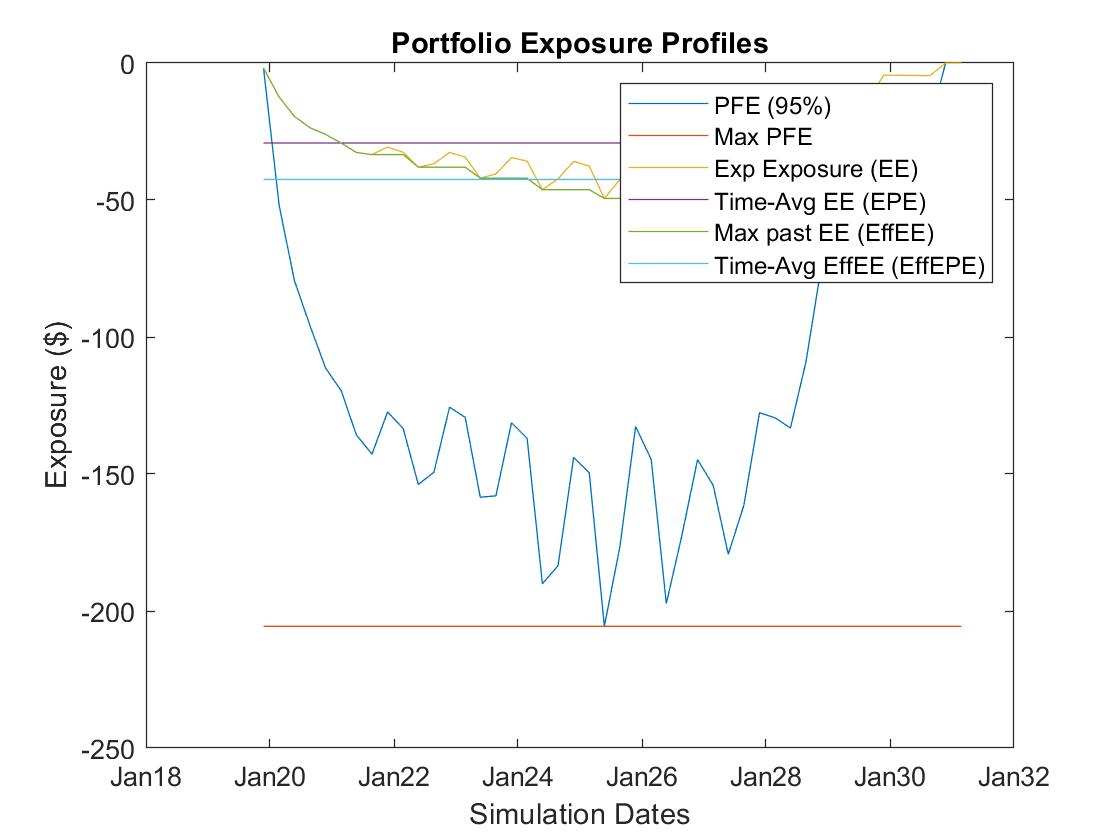}}}
\end{figure}

\begin{figure}[!htb]
\caption{Swaps Portfolio IG FCA Exposure Profiles}
\centerline{\scalebox{0.4}[0.2]{\includegraphics{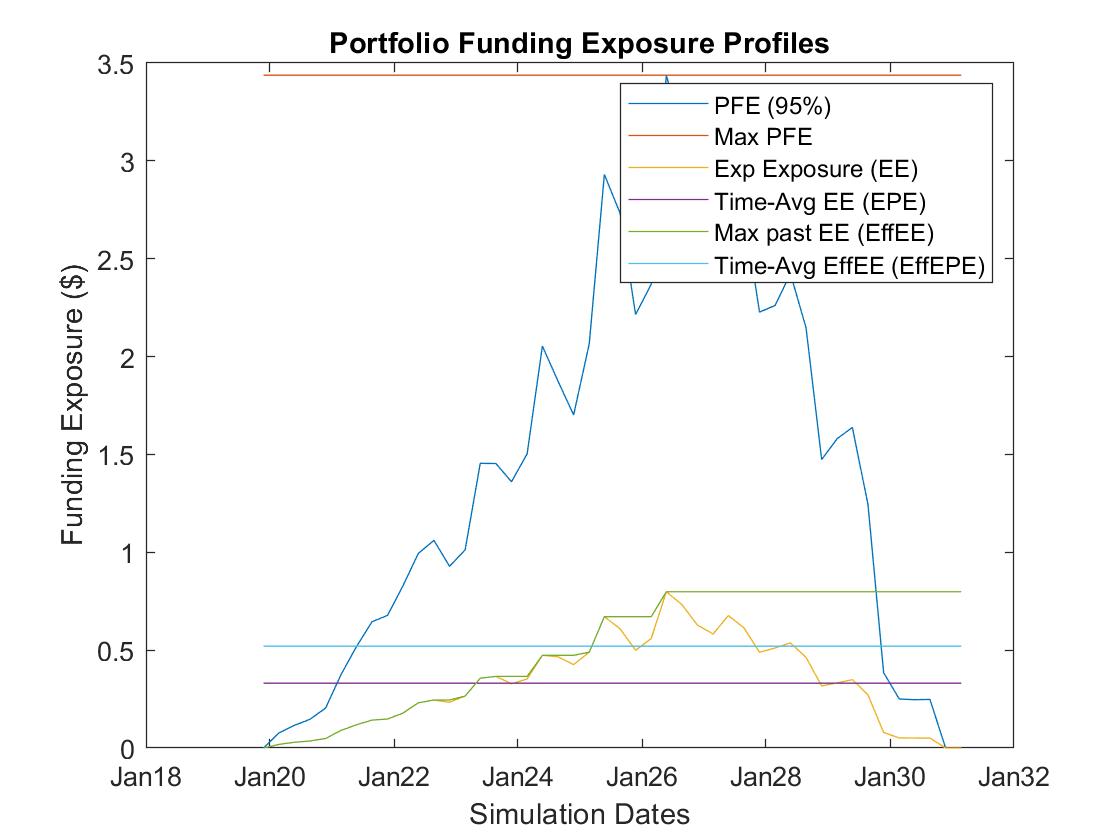}}}
\end{figure}

\begin{figure}[!htb]
\caption{Swaps Portfolio IG FBA Exposure Profiles}
\centerline{\scalebox{0.4}[0.2]{\includegraphics{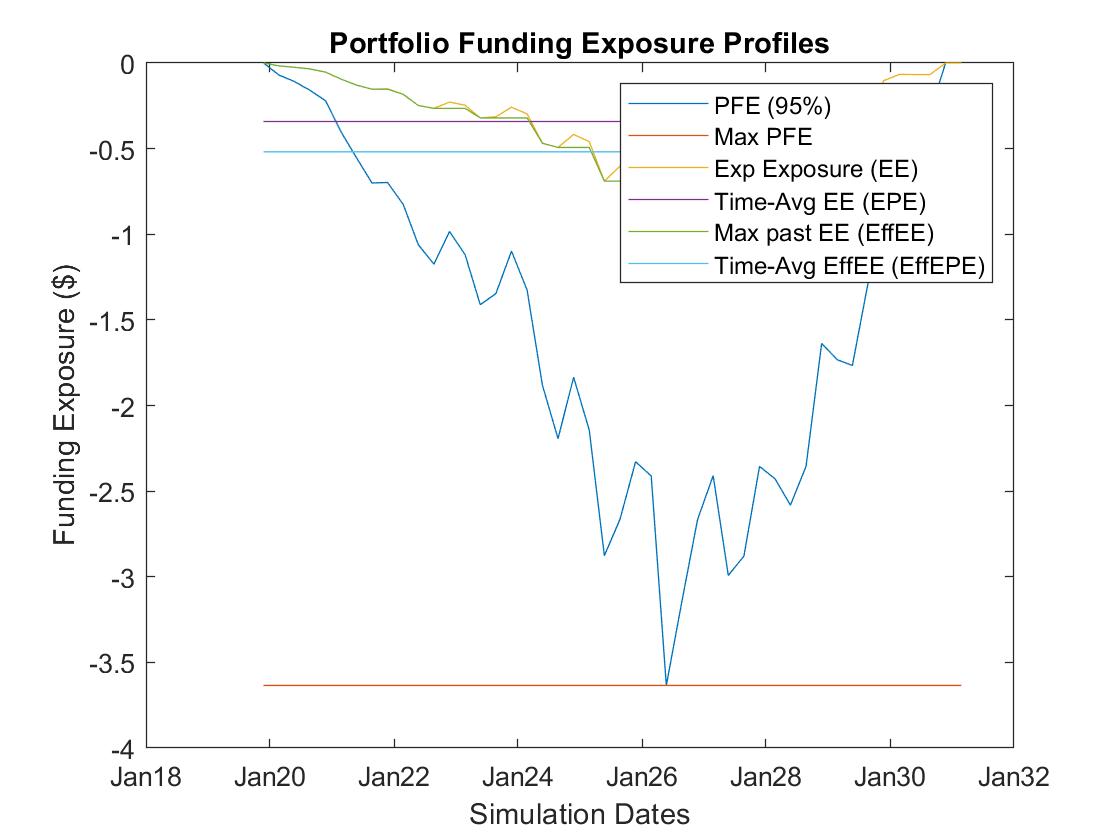}}}
\end{figure}

\begin{figure}[!htb]
\caption{Swaps Portfolio Worst Case IG FVA Profile}
\centerline{\scalebox{0.4}[0.2]{\includegraphics{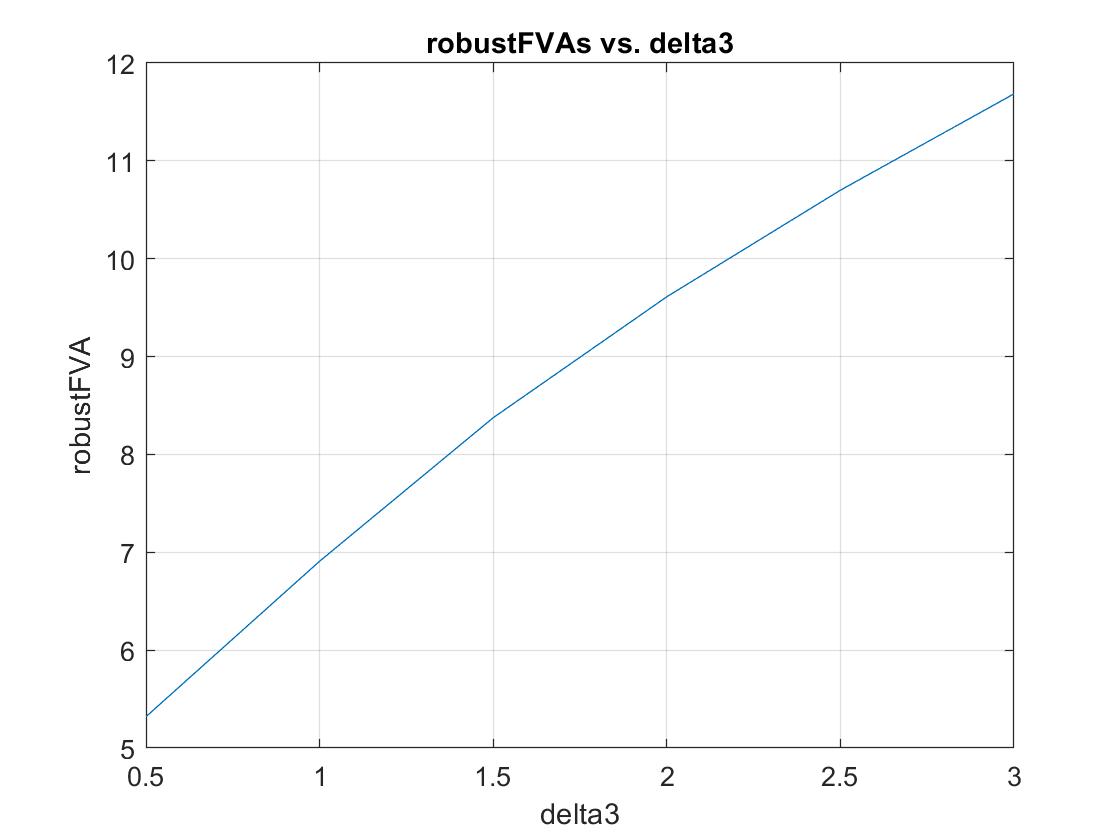}}}
\end{figure}


The baseline FVA for this portfolio is small (less than 1k USD) and represents the dot product of the discounted portfolio FCA exposure profile times joint survival probability plus dot product of the discounted portfolio FBA exposure times joint survival probability. The worst case FVA curve is shown below. Note the worst case FVA is approximately 57\% the size of FCA PFE for Wasserstein radius $\delta$ about 1.9 which maps to a significance level around $95\%$. So the takeaway here is worst case FVA is still a significant percentage of funding PFE for swap portfolios with low counterparty default curves (investment grade). It is also interesting to compare worst case FVA vs. worst case FCA for a given delta. For example, for $\delta$ of 1.7, which represents the $95\%$ significance level for FCA, we see FVA of around 9 which is below FCA of 11.8. This agrees with intuition that FVA should be less than FCA due to funding benefit from FBA.\par


\subsubsection{Portfolio of Interest Rate Swaps, High Yield Counterparty and Firm}

The corresponding FCA portfolio is used for comparison. The portfolio consists of a dozen interest rate swaps, with a mix of receving fixed and paying fixed swaps, at different coupons, maturities, and notionals. The fixed coupons range between 2\% and 2.5\%, the maturities range between 4y and 12y, the notionals range between 400k USD and 1mm USD. The high yield counterparty and firm credit spreads are set to 320 basis points. The table of confidence levels $\beta$ and their corresponding Wasserstein radii $\delta$ follows. \par
\begin{table}[h]
\begin{center}
\caption{FVA High Yield Wasserstein Radii}
\begin{tabular}{ |c|c|c|c|c|c|c| }
 \hline
Confidence Level & 0.80 & 0.85 & 0.90 & 0.95 & 0.99 & 0.999 \\
 \hline
W Radius delta & 3.8 & 4.2 & 4.6 & 5.3 & 6.6 & 8.1 \\
 \hline
\end{tabular}
\end{center}
\end{table}

The scale factor $S_3$ is set (by default) to 1 and the portfolio exposures are scaled to be in units of thousands of dollars. Same comments as above, for FCA and FBA, regarding scaling, apply. Matlab plots characterizing the FVA positive and negative exposure profiles and trajectory of worst case FVA as a function of Wasserstein radius are shown.  \par


\begin{figure}[!htb]
\caption{Swaps Portfolio Positive Exposure Profiles}
\centerline{\scalebox{0.4}[0.2]{\includegraphics{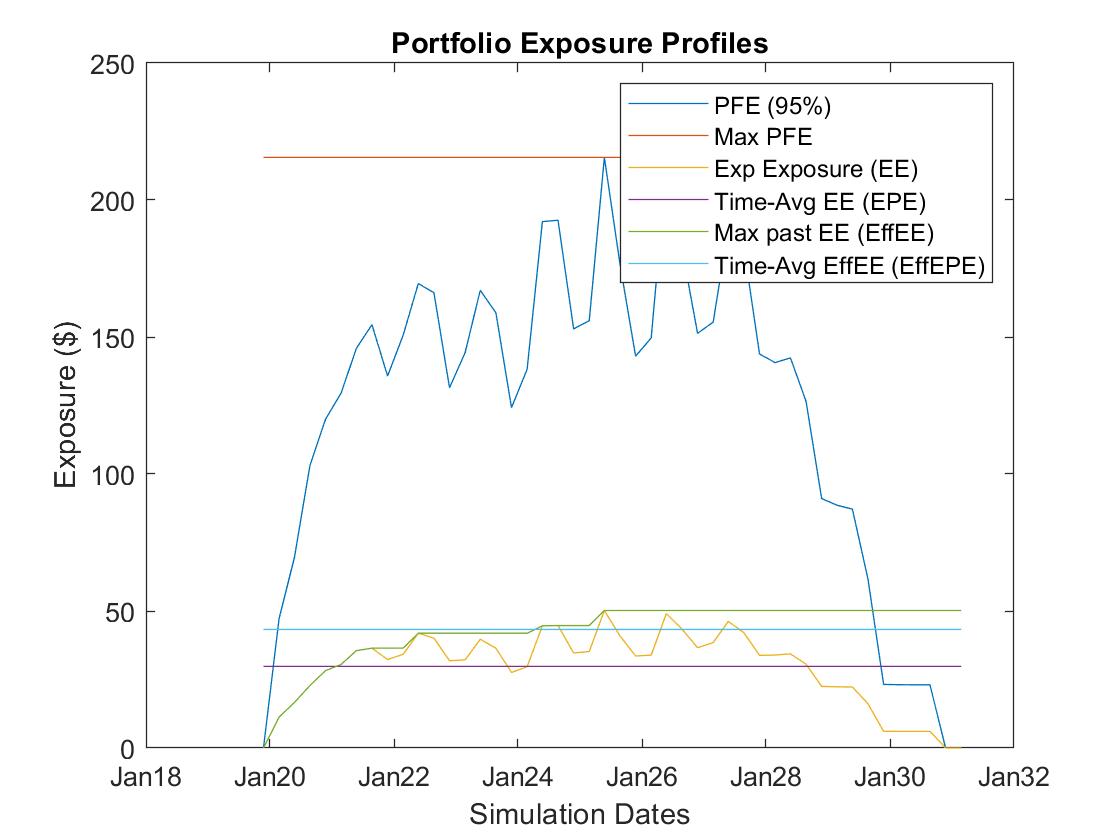}}}
\end{figure}

\begin{figure}[!htb]
\caption{Swaps Portfolio Negative Exposure Profiles}
\centerline{\scalebox{0.4}[0.2]{\includegraphics{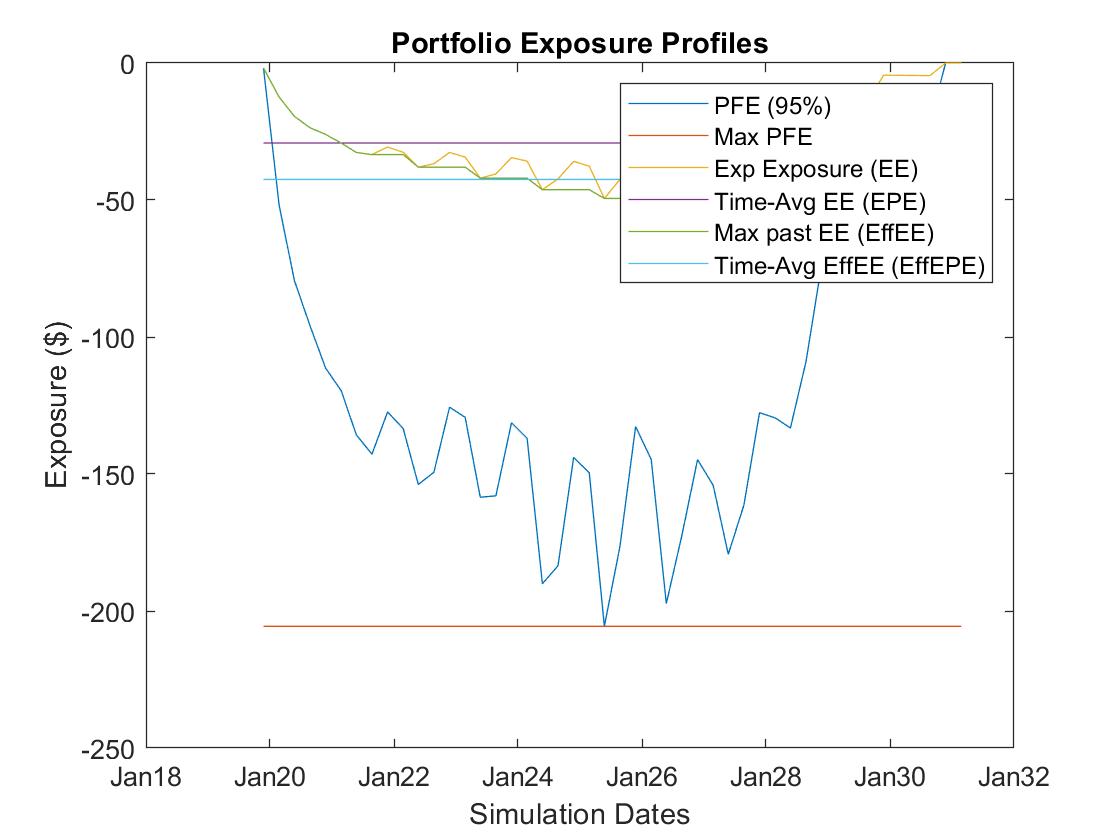}}}
\end{figure}

\begin{figure}[!htb]
\caption{Swaps Portfolio HY FCA Exposure Profiles}
\centerline{\scalebox{0.4}[0.2]{\includegraphics{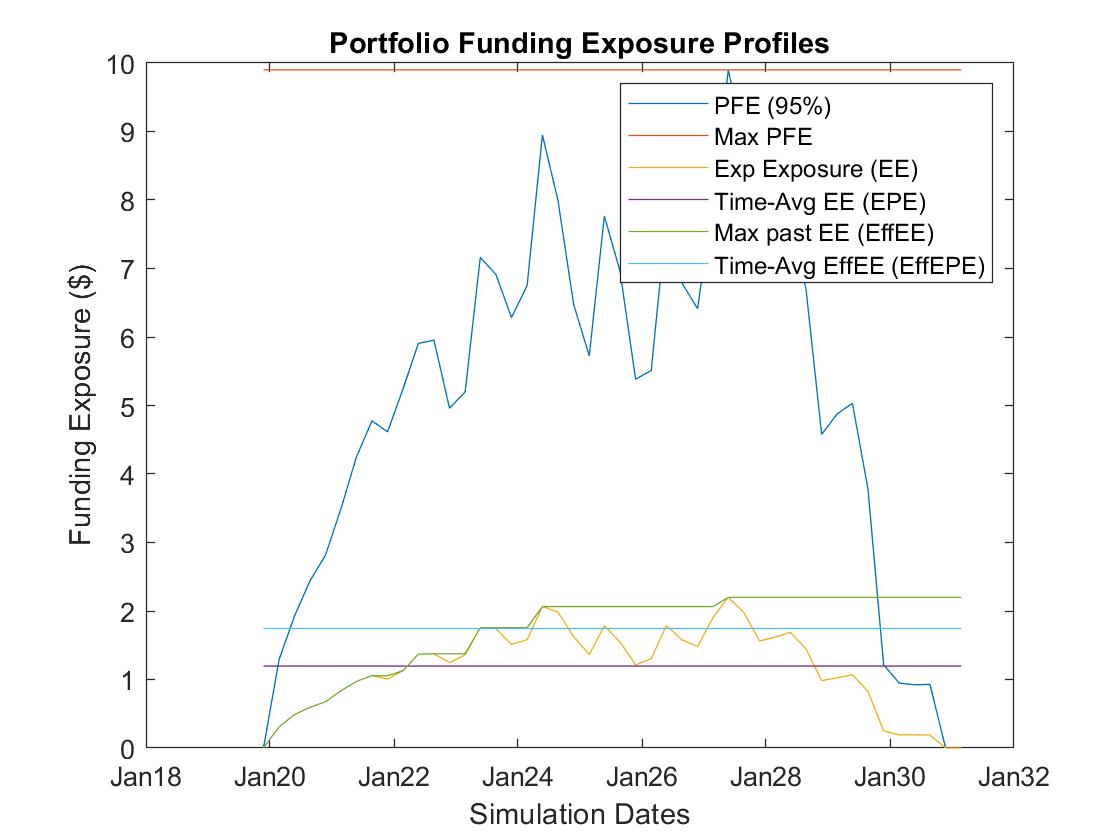}}}
\end{figure}

\begin{figure}[!htb]
\caption{Swaps Portfolio HY FBA Exposure Profiles}
\centerline{\scalebox{0.4}[0.2]{\includegraphics{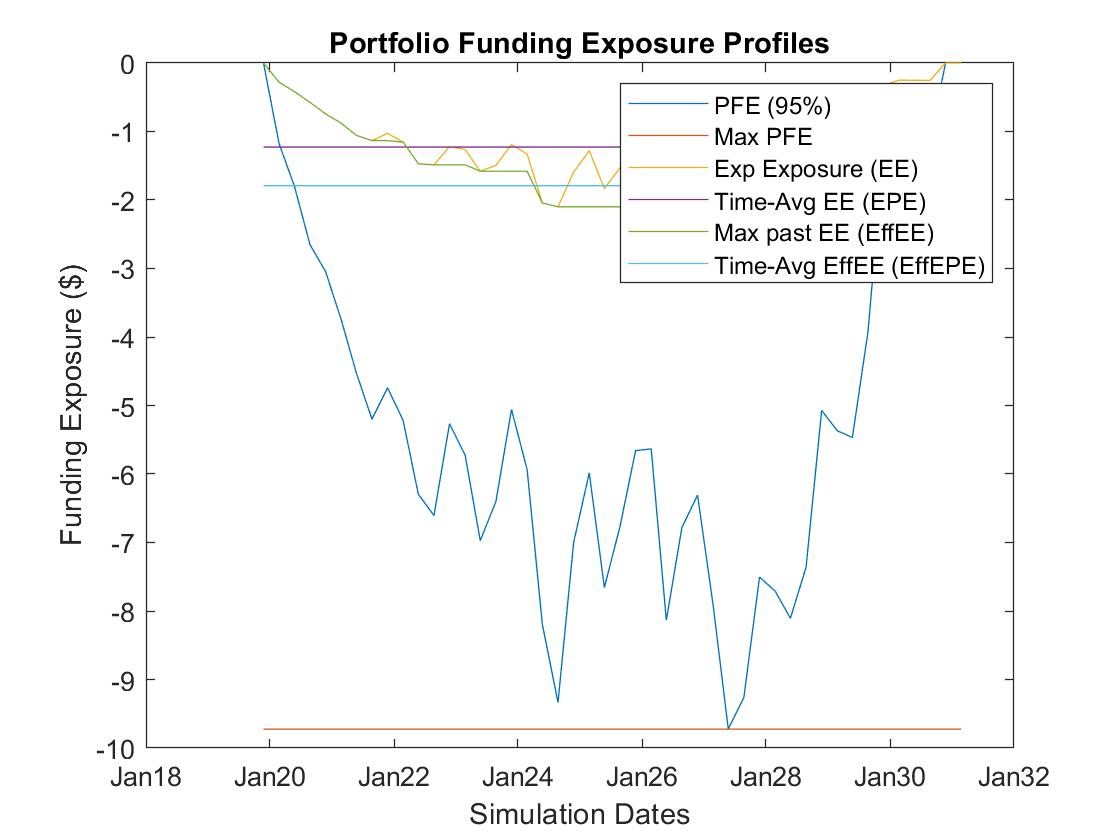}}}
\end{figure}

\begin{figure}[!htb]
\caption{Swaps Portfolio Worst Case HY FVA Profile}
\centerline{\scalebox{0.4}[0.2]{\includegraphics{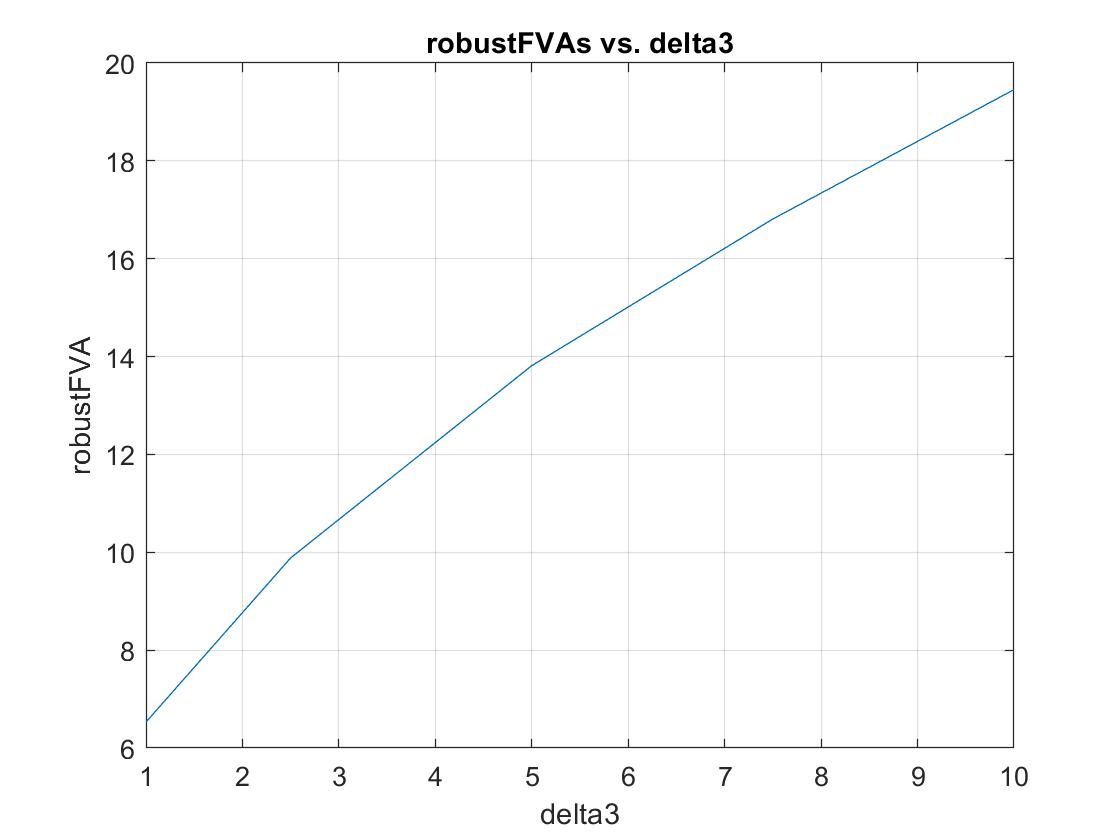}}}
\end{figure}
%
%
The baseline FVA for this portfolio is small (less than 1k USD) and represents the dot product of the discounted portfolio FCA exposure profile times joint survival probability plus dot product of the discounted portfolio FBA exposure times joint survival probability. The worst case FVA curve is shown below. Note the worst case FVA is approximately 23\% the size of FCA PFE for Wasserstein radius $\delta$ about 5.3 which maps to a significance level around $95\%$. So the takeaway here is worst case FVA is still a significant percentage of funding PFE for swap portfolios with high counterparty and firm default curves. It is also interesting to compare worst case FVA vs. worst case FCA for a given delta. For example, for $\delta$ of 4.2, which represents the $95\%$ significance level for FCA, we see FVA of around 12.5 which is below FCA of 19. This agrees with intuition that FVA should be less than FCA due to funding benefit from FBA.\par


\clearpage
\section{Conclusions and Further Work}

This work has developed theoretical results and investigated calculations of robust FVA and wrong way risk for OTC derivatives under distributional uncerainty using Wasserstein distance as an ambiguity measure. The financial market overview and foundational notation and wrong way risk (robust FVA) primal problem definitions were introduced in Section 1. Using recent duality results \citep{blanchetFirst}, the simpler dual formulation and its analytic solutions for FCA, FBA, and FVA were derived in Section 2. After that, in Section 3, some computational experiments were conducted to measure the additional FCA charge (and/or FBA impairment) due to distributional uncertainty for a variety of portfolio and market configurations for FCA, FBA, and FVA. Using some probability results on bounding Wasserstein distance between distributions \citep{Carlsson2018}, a mapping between Wasserstein radii $\delta$ and significance levels $\beta$ was devised to study the trajectories of wrong way risk as a function of radius $\delta$. FCA increased to a significant percentage of PFE. FBA quickly reached its lower bound of zero funding benefit. FVA was below FCA (as expected) but still showed an upward (apparently concave) trajectory as radius $\delta$ increased. Finally, we conclude with some commentary on directions for further research. \par

One direction for future research, as has been previously discussed, is a thorough study (including sensitivity analysis) regarding the pairings of scale factors $(S_1,S_2,S_3)$ and units of portfolio exposures to investigate suitable (unsuitable) ranges that preserve (distort) the shape of the robust FCA, FBA, FVA profiles (as a function of Wasserstein radii, and hence distributional uncertainty) respectively. As a reminder, the intent of scaling is to provide appropriate penalty to the adversarial change in joint distribution of portfolio exposures and default times that promotes worst case FCA, FBA, FVA and wrong way risk. Another direction for future research would be to develop (and apply) similar theoretical machinery as used for robust FVA and wrong way risk in this work towards robust KVA (Capital Valuation Adjustment) and MVA (Margin Valuation Adjustment) and wrong way risk in that context. Intuitively, wrong way risk arises in that context when the market cost of capital and/or funding the margin position increases at the same time as the portfolio exposure increases.\par


\clearpage
\bibliographystyle{apalike}
\bibliography{RobustFVAJ}

\end{document}